\documentstyle[12pt,epsf,epsfig]{article}
\newfont{\thiplo}{msbm10 scaled\magstep 2}
\newfont{\gothic}{eufb10 scaled\magstep 2}
\newfont{\unc}{eurb10}  
\newskip\humongous \humongous=0pt plus 1000pt minus 1000pt
\def\caja{\mathsurround=0pt}
\def\eqalign#1{\,\vcenter{\openup1\jot \caja
        \ialign{\strut \hfil$\displaystyle{##}$&$
        \displaystyle{{}##}$\hfil\crcr#1\crcr}}\,}
\newif\ifdtup

\def\eqright #1\cr{\noalign{\hfill$\displaystyle{{}#1}$}}
\def\eqleft #1\cr{\noalign{\noindent$\displaystyle{{}#1}$\hfill}}

\def\oldreffmt#1{\rlap{[#1]} \hbox to 2\parindent{}}

\def\figfmt#1{\rlap{Figure {#1}} \hbox to 1in{}}

\def\sectioneq{\def\theequation{\thesection.\arabic{equation}}{\let
\holdsection=\section\def\section{\setcounter{equation}{0}\holdsection}}}%

\newcounter{holdequation}

\def\auto{\eqno(\refstepcounter{equation}\theequation)}
\def\begineq #1\endeq{$$ \refstepcounter{equation}\eqalign{#1}\eqno
	(\theequation) $$}
\def\contlimit{\,{\hbox{$\longrightarrow$}\kern-1.8em\lower1ex
\hbox{${\scriptstyle (a\rightarrow0)}$}}\,}
\def\centeron#1#2{{\setbox0=\hbox{#1}\setbox1=\hbox{#2}\ifdim
\wd1>\wd0\kern.5\wd1\kern-.5\wd0\fi
\copy0\kern-.5\wd0\kern-.5\wd1\copy1\ifdim\wd0>\wd1
\kern.5\wd0\kern-.5\wd1\fi}}
\def\centerover#1#2{\centeron{#1}{\setbox0=\hbox{#1}\setbox
1=\hbox{#2}\raise\ht0\hbox{\raise\dp1\hbox{\copy1}}}}
\def\centerunder#1#2{\centeron{#1}{\setbox0=\hbox{#1}\setbox
1=\hbox{#2}\lower\dp0\hbox{\lower\ht1\hbox{\copy1}}}}
\def\lsim{\;\centeron{\raise.35ex\hbox{$<$}}{\lower.65ex\hbox
{$\sim$}}\;}
\def\gsim{\;\centeron{\raise.35ex\hbox{$>$}}{\lower.65ex\hbox
{$\sim$}}\;}
\def\st#1{\centeron{$#1$}{$/$}}

\def\super#1{\ifmmode \hbox{\textsuper{#1}}\else\textsuper{#1}\fi}
\def\textsuper#1{\newcount\holdspacefactor\holdspacefactor=\spacefactor
$^{#1}$\spacefactor=\holdspacefactor}

\def\getcite#1,{\advance\citenumber by1
\def\getcitearg{#1}\def\lastarg{@}
\ifnum\citenumber=1
\ref{#1}\let\next=\getcite\else\ifx\getcitearg\lastarg\let\next=\relax
\else ,\ref{#1}\let\next=\getcite\fi\fi\next}

\def\pom{{\rm P\kern -0.53em\llap I\,}}
\def\spom{{\rm P\kern -0.36em\llap \small I\,}}
\def\sspom{{\rm P\kern -0.33em\llap \footnotesize I\,}}

\relax
\def\contlimit{\,{\hbox{$\longrightarrow$}\kern-1.8em\lower1ex
\hbox{${\scriptstyle (a\rightarrow0)}$}}\,}
\def\upon #1/#2 {{\textstyle{#1\over #2}}}
\relax
\renewcommand{\thefootnote}{\fnsymbol{footnote}} 

\def\mainhead#1{\setcounter{equation}{0}\addtocounter{section}{1}
  \vbox{\begin{center}\large\bf #1\end{center}}\nobreak\par}
\sectioneq

\def\til#1{\centeron{\hbox{$#1$}}{\lower 2ex\hbox{$\char'176$}}}
\def\tild#1{\centeron{\hbox{$\,#1$}}{\lower 2.5ex\hbox{$\char'176$}}}
\def\sumtil{\centeron{\hbox{$\displaystyle\sum$}}{\lower
-1.5ex\hbox{$\widetilde{\phantom{xx}}$}}}

\newcommand{\bit}{\begin{itemize}}
\newcommand{\eit}{\end{itemize}}

\newcommand{\beq}{\begin{equation}}
\newcommand{\eeq}{\end{equation}}
\newcommand{\beqa}{\begin{eqnarray}}
\newcommand{\eeqa}{\end{eqnarray}}

\begin{document} 

\begin{titlepage} 

\rightline{\vbox{\halign{&#\hfil\cr
&ANL-HEP-PR-01-017 \cr
&\today\cr}}} 
\vspace{0.25in} 

\begin{center} 
 
{\large\bf CHIRALITY VIOLATION IN THE }

{\large \bf QCD HIGH-ENERGY S-MATRIX}\footnote{Work 
supported by the U.S.
Department of Energy, Division of High Energy Physics, \newline Contracts
W-31-109-ENG-38 and DEFG05-86-ER-40272} 
\medskip

Alan. R. White\footnote{arw@hep.anl.gov }

\vskip 0.6cm

\centerline{High Energy Physics Division}
\centerline{Argonne National Laboratory}
\centerline{9700 South Cass, Il 60439, USA.}
\vspace{0.5cm}

\end{center}

\begin{abstract} 

In a previous paper it has been shown that the infra-red divergence
associated with the triangle graph axial anomaly 
can occur in triple-regge multi-reggeon
interactions due to unphysical asymptotic triple discontinuities. In 
this paper an asymptotic discontinuity analysis is applied to
high-order feynman diagrams to show that the anomaly exists in contributions
to the triple-regge nine-reggeon interaction. This implies that 
the anomaly occurs in the interactions of 
reggeon states that have the quantum numbers of 
the anomaly current and establishes a direct 
connection with the well-known U(1) problem.

\end{abstract}

\renewcommand{\thefootnote}{\arabic{footnote}} \end{titlepage}

\mainhead{1. INTRODUCTION} 

Perhaps the most important property of a non-abelian gauge theory is the 
existence of non-perturbative euclidean clasical solutions with 
non-trivial topology. If the 
theory is quantized (in principle at least) via the euclidean path-integral,
such solutions produce additional  interactions that,
even if it is not well understood how to evaluate them, are believed to
modify properties of the theory significantly. In particular, 
the topological field configurations produce zero modes
of the Dirac operator which\cite{aj} 
prevent the gauge-invariant separation of 
massless fermion fields into  
chiral (right- and left-handed) components that create 
particles and antiparticles in a well-defined way. As a result,
the number of right- and left-handed Dirac particles is not separately 
conserved. The corresponding U(1) vector charge remains conserved,
but the axial charge conservation that is present in perturbation theory
is violated non-perturbatively, as allowed
by the anomaly in the U(1) axial current. We expect, therefore, that 
quark chirality transitions will play an important role    
within the QCD bound-state S-Matrix. (It is well-known 
that such processes can generate a mass\cite{gth} 
for a bound-state with the
quantum numbers of the $\eta'$.) Clearly, how such transitions contribute
must be a major component of a complete
non-perturbative definition of
the massless theory. 

In this paper we will provide evidence for a completely different 
argument that 
there should be non-conservation of the U(1) axial charge in the 
high-energy (multi-regge)
QCD bound-state S-Matrix. This argument makes no mention
of either the euclidean region, path-integral quantization, or zero modes.
Rather it is part of a program which has gradually taken shape over the 
years\cite{arw99,arw001} and
which we expect to carry out in detail in future papers. The aim of the 
program is to
construct the massless, multiparticle, multi-regge S-Matrix starting from
a spontaneously-broken theory in which gluons and quarks are massive,
and regge behavior and the unitarity
properties of reggeon diagrams
are well-established perturbatively\cite{fkl,jb}.
The massless theory that is our ultimate goal is, of course, 
infra-red divergent and it is widely believed 
that even if the divergences could be handled systematically the large
order behavior of the perturbation
expansion is so bad 
that it can not be used (by itself) to define the theory.
However, we plan to initially apply our construction to the case in which
the gauge coupling does not increase in the infra-red 
region\footnote{This requires an infra-red fixed point in the massless 
$\beta$-function which, in turn, is likely to require a special fermion 
content.}
and the divergence of the perturbation series is reduced considerably.
In addition, the multi-regge region  
may be a special situation. 
Because high-energy is involved, the S-Matrix should be close to
perturbation theory while, because low momentum transfers are also involved, 
$t$-channel unitarity properties
involving the physical spectrum must also be satisfied. 

If it is possible to construct the multi-regge behavior of a 
weak-coupling, massless, 
theory from the essentially perturbative 
starting point of massive reggeon diagrams,
without reference to the low-energy solution of the
theory, then there must be some element that
can produce the ``non-perturbative'' properties of 
confinement and chiral symmetry breaking in the spectrum. We expect that,
in our construction, this will be the triangle anomaly
``chirality violation'' that we argued in \cite{arw99}
can occur\footnote{We will discuss precisely what 
we mean by ``chirality violation'' and ``the anomaly''
in this context shortly.} in certain reggeized gluon interactions due to a
quark loop.
In our future papers, we hope to demonstrate that a multi-regge
S-Matrix with all the desired properties is obtained by 
combining perturbative multi-regge
infra-red behavior with a treatment of the anomaly  
that clearly breaks the U(1) axial symmetry.
As a prelude to the full program, therefore, we must first establish 
that the U(1) anomaly (and also the chiral flavor anomaly - which plays a 
crucial role in the chiral symmetry breaking) does indeed appear in the 
framework of multiparticle multi-regge behavior. Our focus 
in this paper will be on explaining in detail the origin of 
this phenomenon. Our hope is that this explanation will add considerably 
to the arguments that we have already given\cite{arw99}. 

Reggeon diagrams contain reggeized gluon (or quark) propagators,
reggeon interaction vertices, and external couplings to the scattering states,
all of which are gauge invariant, even in perturbation theory. 
In general, many Feynman diagrams
give contributions to a single reggeon diagram and, therefore, to a 
single reggeon interaction. In addition, a reggeon 
interaction vertex has a significance that goes beyond it's perturbative 
description. As we will see, the essential ``non-perturbative'' element of the
calculations we describe is the role played by unphysical asymptotic
multiple discontinuities and the anomaly in contributions to
reggeon interaction vertices that they give. The anomaly in such contributions
can be non-perturbative in the sense that it
can be present in a reggeon vertex but not 
necessarily produce an effect in a perturbative 
amplitude in which it is contained. This is because
the structure of external couplings and additional symmetries of full
reggeon diagrams can be sufficient to produce a cancelation. 
Indeed, we argued in \cite{arw99} that when the scattering states are 
elementary quarks or gluons the anomaly always cancels in the full 
scattering amplitude. Conversely, we expect it to play a crucial role in 
the scattering of ``non-perturbative'' physical bound states.

Before enlarging further on the significance of unphysical asymptotic
multiple discontinuities and the results of this paper,
it will be helpful to briefly discuss the potential relationship 
between our work and the euclidean path-integral formalism. We note, first, 
that in Minkowski space the Dirac zero modes due to topological gauge fields 
are manifest\cite{aj} as the spectral flow of the eigenvalues of
the corresponding (gauge-dependent) ``Hamiltonian''. Since there 
is no complete non-perturbative Hamiltonian formalism for QCD, 
there is no well-developed understanding of what the general  
consequences of spectral flow might be. The common expectation is probably
that such phenomena will be overwhelmed
by strong-coupling effects of the kind usually assumed to be associated with
confinement. However, in a massless theory in which the gauge coupling remains 
small in the infra-red region, we should not expect this to be the case.
As a minimum, we anticipate that (in an appropriate background field)
zero energy fermion states identified initially 
as a particle (presumably within a boundstate) can evolve with time into 
a filled vacuum state of the corresponding Dirac sea 
and, similarly, filled vacuum states can evolve into particles. (The existence
of stable bound states and physical scattering processes in such an
environment is surely far from trivial!)

The U(1) anomaly can be interpreted in terms of infra-red
spectral flow as follows.
Associated with the anomalous divergence equation, 
the massless axial-vector triangle graph has\cite{cg} an infra-red divergence  
that involves a  zero four-momentum
propagator. Both the ``particle'' and ``antiparticle'' poles  
of the propagator contribute to the 
divergence. The coupling at one end of the propagator can be viewed 
as the vertex for production of the particle while simultaneously (and
symmetrically) that
at the other end describes the production of the antiparticle.
For the propagator to describe a physical zero momentum 
transition there must be 
spectral flow (due to a background gauge field) 
in that the production of the antiparticle (or the particle)
must be counted as the absorption of a particle (antiparticle). In this case, 
the transition becomes a ``chirality transition''.  
Consequently, in Minkowski space the U(1) divergence equation  
provides a connection between the topological 
structure of a background gauge field and the net 
infra-red ``spectral flow'' of zero momentum,
massless, Dirac particle and antiparticle states. 

In our analysis ``spectral flow'' is introduced by the appearance
of the triangle graph infra-red divergence (referred to above as 
the ``anomaly'') in reggeized gluon interactions.  
The triangle graph appears as an effective interaction generated by
particular multi-gluon interactions due to a quark loop. 
There are no axial-vector 
currents in the QCD interaction but, as we already described in \cite{arw99},
multi-regge effective interactions can contain  
components of an axial-vector interaction. In sufficiently high order,
interactions appear involving reggeon states with the quantum 
numbers of the anomaly (winding-number) current. Remarkably, perhaps,
we will establish in 
this paper that it is interactions of this last 
kind that have the contributions from unphysical multiple
discontinuities that we argued in \cite{arw99} are necessary for the
anomaly infra-red divergence to appear. As a result, the
U(1) problem is clearly encountered.
Indeed, we anticipate that the infra-red discussion we give 
is connected to ``ultra-violet'' problems (involving momenta flowing 
around the internal quark loop that are comparable in magnitude to 
the large external momenta) that reflect 
the usual relationship between infra-red and ultra-violet 
manifestations of the anomaly. We would expect  
short-distance interactions of the winding number current to appear
directly in this ultra-violet context. (It is, perhaps, 
unfortunate that the anomaly 
is a high-order, many gluon, phenomenon. However,
this is to be expected if the anomaly current, containing a 
product of three gluon fields, has to be  involved.)

We will call the basic process, in which a physical region
zero momentum propagator contributes to a triangle graph divergence, 
a chirality transition and will refer 
to the general phenomenon as 
``chirality violation'', although we could
equally well call it spectral flow\footnote{Neither description is strictly
appropriate. Since we study only S-Matrix 
elements we can not define chirality via right and left-handed fields and 
since we do not have a hamiltonian we also can not define 
spectral flow in the normal manner. We also can not define the anomaly in 
terms of the divergence of an
axial current although we can, as we discussed in \cite{arw99},
relate it to the violation of reggeon Ward identities that normally are 
a consequence of gauge invariance.}. It will also be what we generically
refer to as ``the anomaly'', within our formalism.
In this paper, as in \cite{arw99},
we will concentrate on the feynman diagram amplitudes that
produce the anomaly in reggeon interactions 
and, apart from the brief description at the end of this Introduction, 
will not discuss the general program any further. 
In \cite{arw99} we distinguished
two methods for calculating multi-regge amplitudes - the direct calculation
of diagrams in light-cone co-ordinates and the calculation of multiple
asymptotic discontinuities with the subsequent use of an asymptotic dispersion
relation. We emphasized that the direct calculational method 
is impractical for the problem we are discussing. This is because of the 
large number of diagrams that could contribute
and because the complexity of the diagrams
makes a full discussion of 
whether or not integration contours are truly trapped, in the asymptotic  
limits involved, very difficult. Consequently the asymptotic dispersion 
relation method has to be used.

The form of the asymptotic dispersion relation for a given multi-regge 
process is determined by the  
asymptotic multiple discontinuities that satisfy the Steinmann relation
property that the discontinuities occur in non-overlapping invariant channels.
Such discontinuities are explicitly reflected in the analytic structure of 
asymptotic amplitudes and, conversely, 
using the dispersion relation, 
amplitudes can be calculated directly 
from the discontinuities\cite{jb,arw00}. 
The crucial feature of the high-order amplitudes that produce reggeon 
interactions containing the anomaly
is the presence of unphysical  multiple discontinuities that 
satisfy the Steinmann relation property.
Such discontinuities are present only in 
complex (imaginary momentum) parts of the asymptotic region for 
more complicated many-particle multi-regge 
processes, the simplest of which is the full triple-regge region\cite{gw}.
However, just because they
are in non-overlapping channels these discontinuities can 
(and must) consistently 
appear in the asymptotic amplitudes that describe also the 
real physical region behavior. 

The familiar amplitudes that appear in multi-regge
production processes (such as 
those that contribute to the BFKL equation\cite{fkl}) 
do not contain unphysical multiple discontinuities.
Rather they contain only multiple discontinuities
that are naturally interpreted as due to a succession of physical
region on-shell scattering processes\cite{jb}. (The necessity for a
physical time-ordering of such processes then determines the absence of
overlapping channel discontinuities.) Because the physical region multiple
discontinuities involve only physical amplitudes and physical intermediate
states, when they are calculated using the perturbative amplitudes
of the massless theory, they can not contain chirality transitions
associated with particle/antiparticle ambiguities. Therefore,
when only production processes are 
involved (i.e. at what we might call the BFKL level of multi-regge theory) 
there is no possibility for ``chirality violation''. 
In more elaborate scattering processes
the unphysical multiple discontinuities appear and they may, a-priori, 
contain potential chirality
transitions, even when calculated perturbatively. This is because
the discontinuities involved may contain what, in a physical region,
would be mixed combinations of forward-going particles and backward
going antiparticles (``mixed $\alpha$ singularities'', in the language
of S-Matrix Theory). Our purpose, in this paper, is to show that this 
phenomenon does occur 
and to demonstrate that, in the massless limit, the reggeon interactions
associated with the asymptotic amplitudes can contain 
the anomaly and so, potentially, can 
produce physical region infra-red divergences.

In our previous paper studying triple-regge interactions\cite{arw99} 
we already discussed why the anomaly could only appear in reggeized 
amplitudes containing unphysical triple discontinuities.
In practise, however, we only studied (what appeared to be) 
the lowest-order relevant  
diagrams, i.e. those that contain two gluons in each $t$-channel. 
We isolated the physical momentum configuration 
within ``maximally non-planar'' 
diagrams that, in the massless (quark)
limit, could potentially give an infra-red 
divergence associated with the anomaly, provided the appropriate 
on-shell propagators contribute to the asymptotic behavior. However,
although they are maximally non-planar these diagrams do not 
have the complexity required to  
contain the unphysical triple discontinuities that, according to our method
of analysis, would determine that these propagators do contribute.
We noted, nevertheless, that the necessary 
discontinuities did appear to be present in
the higher-order amplitudes that would give the reggeization of the gluons 
in the diagrams we studied. Therefore, we argued, the anomaly
configurations in the lowest-order diagrams
could  be required as (generalized) real parts
needed to accompany the higher-order unphysical
triple discontinuities. Paradoxically, perhaps,
we simultaneously suggested that there would 
be cancelations among diagrams 
such that the anomaly would survive only when reggeon states
with the quantum numbers of the winding-number current are involved. 

In this paper we will study
the high-order reggeization diagrams in detail and will find that the 
situation is actually simpler than we suggested. The anomaly infra-red
divergence is produced by a quark loop in which many propagators 
are on-shell and one 
propagator carries the zero momentum and energy that allows a chirality
transition. The on-shell conditions have to be associated with a 
triple discontinuity in such a way that a triplet
of the on-shell particles (each associated with a separate discontinuity) 
are all quarks (or all antiquarks). 
It is straightforward to see that this ``all quarks'' requirement
can not be satisfied by a physical discontinuity and that, in fact, it is 
very difficult to satisfy. Indeed, we find that 
the reggeization diagrams that we 
suggested in \cite{arw99} might contain the anomaly actually 
do not satisfy the all-quarks requirement. As we proceed to higher orders
we eventually find that this requirement is satisfied. However, a final
requirement that the spin structure that generates the anomaly also be 
present, further restricts the triple discontinuities that can
contribute. Eventually
we arrive at (a small class of) diagrams
that contain a triple discontinuity with all the right properties. 
However, this discontinuity is truly unphysical in that it 
occurs as a combination of three ``asymptotic pseudothresholds'' 
each of which contains particles, effectively,
going in opposite time directions. 
The reggeon interactions produced are also of sufficiently high order that 
the minimum circumstances in which they can occur (between color zero
reggeon states) is when each of the states involved carries
the quantum numbers of the U(1) anomaly current. Nevertheless, 
this establishes the essential result of this paper that the 
triangle anomaly does 
occur in reggeized gluon interactions extracted from unphysical multiple 
discontinuities and that the phenomenon we are
discussing is indeed the U(1) anomaly. The results of this paper
also imply that that the lower-order diagrams considered in \cite{arw99},
although valuable to discuss for illustrative processes, are 
essentially irrelevant. 

In this paper we are satisfied to simply demonstrate
that there are diagrams which generate a reggeon interaction 
in which the anomaly appears, and that the reggeon
states involved have the quantum numbers of the anomaly current. We
do not discuss whether there are cancelations that could occur.
We postpone this for the following papers that will lay out the details 
of the construction of the bound-state S-Matrix alluded to above.
For the moment we note only that triple-regge interactions of
the kind we consider here will contribute generally to the vertices 
and interactions of the reggeon bound states that emerge 
and refer to the brief discussion 
in \cite{arw99}, and also the outline in
\cite{arw001}, for more details. A brief, general,
description of the anticipated construction is as follows. 

We expect to obtain massless QCD from the massive theory 
in two stages. In the first stage, the  
(spontaneously-broken) gauge symmetry is restored to SU(2). 
The U(1) chiral symmetry is broken by the introduction of a ``wee-parton''  
condensate with anomaly current quantum numbers in 
scattering reggeon bound states.
Our expectation is that an anomaly infra-red divergence then appears
and determines the ``physical scattering amplitudes''. After the divergence
is factorized off, the condensate self-consistently
appears in all intermediate and final reggeon states.    
An essential ingredient will be to show\cite{arw99}
that if we regulate the anomaly ultra-violet divergences involved, 
the infra-red divergence gives a result that is independent of the
regularization used. (Note that,
by identifying, regulating, and organizing how chirality violation 
divergences produce reggeon states and scattering processes,
we constrain how, in the path-integral formalism, 
non-perturbative topological gauge fields must contribute to the 
massless theory. Indeed, if we succeed in our goals,
we will implicitly determine how the  
spectral flow of the Dirac sea must contribute if 
a unitary high-energy S-Matrix is to be obtained.)

U(1) chirality violation appears within interactions of
the pomeron and additional chirality violation, related to the 
anomaly in flavor current vertices, is responsible for the appearance 
of the pion and the ``nucleon''. (At this stage, 
the nucleon (to be) is a chiral Goldstone boson, just like the pion.)
The pomeron should be in a supercritical phase
of Reggeon Field Theory and the spectrum of 
bound-states should have both SU(2) confinement
and chiral symmetry breaking\footnote{It is possible, if not likely, 
that the role of the
Dirac sea in producing confinement in this context is related to that 
proposed by Gribov\cite{gr}.}. In the second stage,
the full gauge symmetry is to be restored
by the randomization of the SU(2) condensate within SU(3). The randomization
should correspond to a phase-transition within Reggeon Field Theory. 
We expect that the 
asymptotic freedom requirement that contact with perturbation theory
remain at short distances determines that the theory must be right 
at the critical point
associated with critical behavior for the pomeron\cite{cri}.

Finally, we note that
while we have not studied the issue in any detail, we believe that 
when the fermions involved are massive only the ``ultra-violet anomaly'' 
is present. As we remarked above, this will produce 
problems (for bound-state scattering amplitudes) 
in the  (internal) momentum region of 
reggeon interactions where the momenta are  
of the order of, or larger, than the external regge limit momenta and effective
interactions of the anomaly current should be 
directly involved. A-priori, we expect the presence of the anomaly, 
in this form,
to lead to the violation of reggeon Ward identities and to 
increased power behavior asymptotically for 
the amplitudes in which it is contained. A consequent 
violation of unitarity bounds by potential bound-state amplitudes
is therefore threatened and this is clearly 
where non-perturbative topological contributions
could be crucial. Of course, if the anomaly produces infra-red divergent
amplitudes in the massless theory, it could also imply that the 
reggeon diagram result for the asymptotic behavior is wrong and unitarity
bounds could be threatened. However, the implication of the infra-red 
divergence structure we envision is that the divergences both select
the physical states and can be absorbed into a redefinition of the
states that leaves reggeon asymptotic behavior intact.  
Thus, as a matter of both principle and practicality we believe
that the massless multi-regge amplitudes must be constructed first - by a 
procedure that regularizes the ultra-violet anomaly and allows the infra-red
behavior to dominate.  
We also believe that it is the relative simplicity of the infra-red structure 
of the anomaly, together with the unitarity properties of reggeon 
diagams and their relationship with Reggeon Field Theory, that will 
allow us to carry out such a procedure.

\newpage

\mainhead{2. MULTIPLE DISCONTINUITIES AND THE STEINMANN RELATIONS.} 

The Steinmann relations originated in axiomatic field theory\cite{ste}. They 
(essentially) describe the restrictions that the time-ordering of interactions
places on the combinations of intermediate states that can occur in a
scattering process. For on-shell S-Matrix amplitudes their significance is
most immediately appreciated in the approximation that we ignore higher-order
Landau singularities and consider only the normal threshold branch points (and
stable particle poles) that occur in individual channel invariants. The
Steinmann relations then say that simultaneous thresholds (and/or poles) can
not occur in overlapping channels. (Channels overlap if they contain a common
subset of external particles.) As a result an $N$-point amplitude has at most
$N-3$ simultaneous cuts (or poles) 
in distinct invariants. The possible combinations of
cuts can be described by tree diagrams with three-point vertices in which each
internal line corresponds to a channel invariant in which there is a cut due
to intermediate state thresholds - as illustrated in Fig.~2.1 for the 7-point
amplitude.
\begin{center}
\epsfxsize=4in
\epsffile{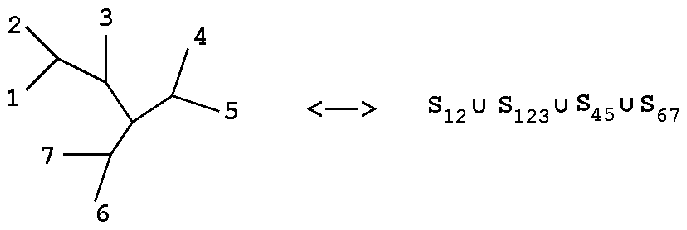}

Fig.~2.1 A Tree Diagram Representing Simultaneous Invariant Cuts.
\end{center}
(As usual, $s_{12}=(P_1+P_2)^2~, s_{123}=(P_1+P_2+P_3)^2~,$
etc.) The set of all combinations of thresholds (and poles) allowed by the 
Steinmann
relations is the basic singularity structure of all scattering amplitudes. The
higher-order Landau singularities are believed\cite{arw00} 
to emerge from the normal
thresholds in a manner that for many purposes (including ours, as we discuss
shortly,) makes them a secondary effect.

Conversely, the combination of cuts represented 
by a particular tree diagram can
be directly associated with a set of physical scattering processes. As
illustrated in Fig.~2.2, this is the set of all processes (involving all the
external particles of the diagram as either ingoing or outgoing particles) in
which it is kinematically possible for all of the internal lines to be
replaced by physical 
multiparticle states\footnote{We do not distinguish processes in
which ingoing and outgoing particles are interchanged via CPT conjugation}.
\begin{center}
\epsfxsize=6in
\epsffile{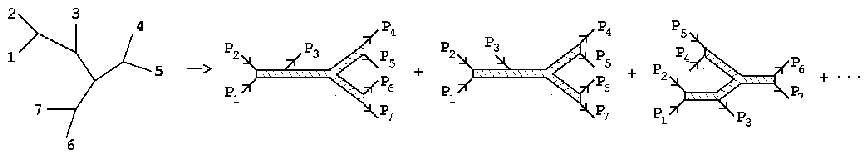}

Fig.~2.2 Physical Scattering Processes Corresponding to Fig.~2.1.
\end{center}
The hatched segments represent physical intermediate states that, if they are 
all placed on shell, give (essentially) the associated multiple discontinuity.

The Steinmann relations play a fundamental role in multi-regge theory. It is 
possible to show\cite{arw00} that in a physical multi-regge asymptotic region 
the analytic structure of scattering amplitudes can be treated as if only 
normal thresholds satisfying the Steinmann relations were present. In effect,
higher-order Landau singularities are suppressed. This has the very important
consequence that only the normal threshold cuts in individual channel 
invariants need be represented by multi-regge
asymptotic formulae. Furthermore, if we consider only the multi-regge limits
accessible in $2 \to M$ production processes, it can be shown that the maximal
number (M-1) of simultaneous thresholds is encountered asymptotically only in
physical regions. This is a generalization of the cut-plane analyticity
property familiar from elastic scattering.

If we consider the multi-regge regions of $M \to M'$ scattering amplitudes
($M,M' \geq 3$) there is a significant change. To understand the point 
involved consider the simplest case of the tree diagram of
Fig.~2.3. At first sight this diagram corresponds only to the 
$2 \to 4$ production processes shown. 
\begin{center}
\epsfxsize=5in
\epsffile{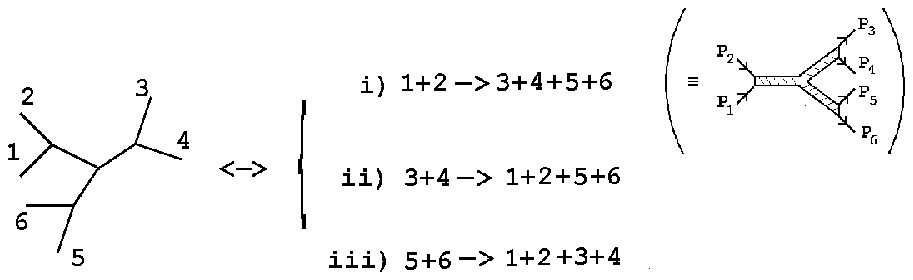}

Fig.~2.3 A Tree Diagram and Corresponding Physical Scattering Processes.
\end{center}
The three distinct scattering processes are distinguished 
by different constraints on the invariants, i.e.
$$
\eqalign{\hbox{i)}& ~\sqrt{s_{12}} ~> ~\sqrt{s_{34}}+\sqrt{s_{56}}~,~~ \cr
\hbox{ii)}&~ \sqrt{s_{34}} ~> ~\sqrt{s_{12}}+\sqrt{s_{56}}~,~~\cr
\hbox{iii)}&~ \sqrt{s_{56}} ~>~ \sqrt{s_{12}}+\sqrt{s_{34}} }
\auto\label{mag}
$$
We can also regard the three processes involved as distinguished by the 
selection of one
pair of particles as incoming, which then must have energy larger than the
sum of the subenergies of the other two pairs, which are necessarily in the 
outgoing state.

We may wonder about the symmetric asymptotic region in which
$$
\sqrt{s_{12}} ~\sim ~\sqrt{s_{34}}~\sim~ \sqrt{s_{56}} ~~\to~~ \infty
\auto\label{mag1}
$$
There are no physical scattering processes in this region. However,
the three processes of (\ref{mag}) are described by the same (analytically
continued) amplitude and so analytic continuation from each of the 
physical regions implies that such cuts must be present. 
It is, perhaps, natural 
that a triple discontinuity should exist that is 
symmetric with respect to the three processes of Fig.~2.3. Apparently, though,
the symmetry requirement could only be satisfied if all the external
particles are in the 
final, or initial, state. However, as we discuss further 
in Sections 4 and 5, 
if we allow particles to carry complex momenta, a positive 
value for a two-particle energy invariant can  
be achieved by a combination of an ``incoming'' and an 
``outgoing'' particle in that they carry opposite sign, but imaginary, 
energies. Therefore, in the symmetric region it is possible for the 
three cuts of Fig.~2.3 
to be present if each is associated with such a combination.    
We will show in the following that
there are unphysical processes (with imaginary momenta) in this region that
do produce a triple discontinuity of this kind and we will refer to it as an 
``unphysical triple disconinuity''.

Because the external particles for each cut are both  
ingoing and outgoing, intermediate states can also be produced that involve
such combinations. As a result, the triple discontinuity can contain the 
``particle - antiparticle'' transitions that ultimately provide  
the massless chirality transitions that we are looking for. 
Moreover, since the complex momentum
part of (\ref{mag1}) is contained in the triple-regge
asymptotic region, the triple discontinuity 
must be present in triple-regge asymptotic formulae.
This is possible just because this 
combination of cuts has the Steinmann property. 
Moreover, because of the potential for chirality violation,
it is a natural context in which to see the anomaly appear.
The importance of the triple-regge region is that it is the simplest
multi-regge limit in which the vertices appear that provide the couplings
of bound-state regge
poles such as the pomeron or the pion. For higher-point $M \to M'$
amplitudes there is a wide range of unphysical multiple discontinuities 
satisfying the Steinmann relations. Bound-state scattering amplitudes can thus
appear in which the anomaly is a crucial element.

\newpage

\mainhead{3. THE PHYSICAL REGION ANOMALY AND THE 
TRIPLE-REGGE DISPERSION RELATION}

In our previous paper\cite{arw99} we studied the full triple-regge
limit\cite{gw} of 
three-to-three quark scattering. 
If we denote the initial momenta as $P_i~, ~i=1,2,3$, and the final momenta 
as $- P_{i'} = P_i + Q_i, ~i=1,2,3$,
the triple-regge limit can be realized, within the physical region, by taking 
each of $P_1,~P_2$ and $P_3$ large 
along distinct light-cones, with the momentum transfers $Q_1, Q_2$ and $Q_3$
kept finite, i.e.
\newline \parbox{3.1in}{ 
$$
\eqalign{ P_1~\to&~ P_{1^+}~= ~(p_1,p_1,0,0)~,~~p_1 \to \infty \cr
P_2~\to&~ {P_2^+}~= ~(p_2,0,p_2,0)~,~~p_2 \to \infty \cr
P_3~\to&~ {P_3^+}~= ~(p_3,0,0,p_3)~,~~p_3 \to \infty  }
$$}
\parbox{2.9in}{
$$ \eqalign{
~~~q_1=Q_1/2~\to&~~ (\hat{q}_1,\hat{q}_1,q_{12},q_{13})\cr
~~~q_2=Q_2/2~\to&~ ~(\hat{q}_2,q_{21},\hat{q}_2,q_{23})\cr
~~~q_3=Q_3/2~\to&~~(\hat{q}_3,q_{31},q_{32},\hat{q}_3)}
\auto\label{np3}
$$}
Momentum conservation gives 
a total of five independent $q$ variables which, along  
with $p_1, p_2$ and $p_3$, give the necessary eight variables. The definition
of the triple-regge limit in terms of angular variables is 
given in \cite{arw99}. For our present purposes the above definition in 
terms of momenta will be sufficient. This will alow us to avoid the
extra complication of defining helicity angles, helicity-pole limits
etc. It will be important that the
asymptotic behavior involved must hold also for all complex values of the large momenta,
including the additional physical regions reached by reversing the 
signs of the $p_i$.

In \cite{arw99} we also studied feynman diagrams that contain a 
closed quark loop and generate
triple-regge reggeized gluon interactions containing the loop.
We considered the lowest-order amplitudes in which the anomaly could 
potentially appear and, in particular, studied maximally non-planar 
diagrams of the 
kind shown in Fig.~3.1(a). 
\begin{center}
\leavevmode
\epsfxsize=4.5in
\epsffile{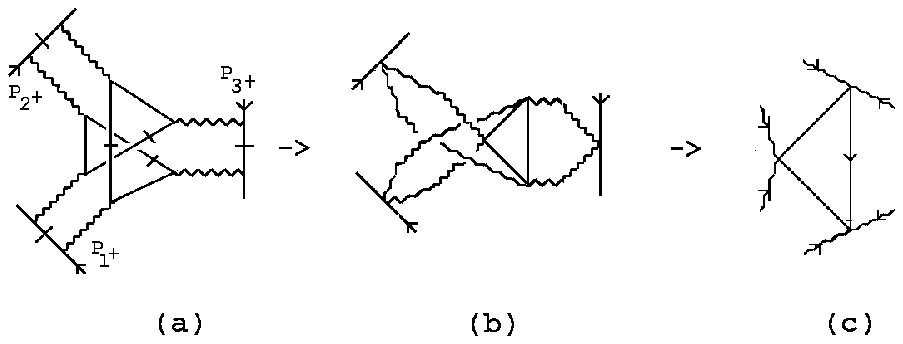}

Fig.~3.1 A maximally non-planar diagram and 
rhe triangle diagram reggeon interaction prouced.   

\end{center}
(As usual, the solid and wavy lines 
respectively represent a quark and a gluon.
We have reversed the direction of $P_3$ relative 
to the notation of \cite{arw99} in order to have a completely symmetric
notation.) The leading asymptotic contributions come from 
regions of gluon loop integrations where some of the 
propagators in the quark loop and the scattering quark systems
are on-shell. We discuss the determination of which propagators can be
on-shell below. For the moment we consider the possibility, discussed at 
length in \cite{arw99}, that the 
on-shell lines are those that are hatched in Fig.~3.1(a). We will eventually 
conclude that this combination of on-shell propagators can not produce a 
reggeon interaction with a physical region anomaly divergence, 
even though it does produce a triangle
diagram interaction. As we will see, the issue is not just which 
propagators are placed on-shell but also which pole (``particle'' or 
``antiparticle'') is involved. (As the discussion in the previous 
Section suggested,
for the unphysical discontinuities, with which we will ultimately 
be concerned, the answer to this question is not necessarily unambiguous.)
In the following we initially 
ignore this subtlety. As it emerges in 
our discussion it will become clear that it is a vital part 
of the search for further diagrams which do produce an interaction 
containing the anomaly.

If the hatched on-shell propagators are used to carry out
light-like longitudinal momentum integrations the integrals over gluon loop
momenta reduce to two-dimensional ``transverse'' integrals over
spacelike momenta, as illustrated by Fig.~3.1(b). The transverse 
plane (and orthogonal light-like momenta)
can, in general, be chosen differently in each $t$-channel. 
If $Q_{i\perp}$ is the projection of $Q_i$ on the corresponding 
transverse plane, the leading asymptotic contribution then has the form
$$
\eqalign{ ~~~~~P_{1^+}~ P_{2^+}~ P_{3^+}~
\prod_{i=1}^3 \int & { d^2 k_{i1}d^2 k_{i2}\over  k_{ i1}^2  k_{i2}^2}  
~~ \delta^2 (Q_{i\perp} -  k_{i1} -  k_{i2})~G^2_i(k_{i1},k_{i2},\cdots)
\cr &~~~~~~~~\times ~ R^6(Q_1,Q_2,Q_3,
k_{11}, k_{12}, \cdots )} \auto \label{211}
$$
where $ R^6(Q_1,Q_2,Q_3,k_{11}, k_{12}, \cdots )$  
can be identified with the ``reduced'',
or ``contracted'', feynman diagram of Fig.~3.1(c). If we write
$$
k_{i1} ~= ~q_i + k_i~, ~~~~ k_{i2} ~= ~q_i - k_i~,
\auto\label{dki}
$$
then (with a particular choice\cite{arw99} of transverse planes)
$$
\eqalign{ &R^6(q_1,q_2,q_3,k_1,k_2,k_3) ~=\cr
& \int d^4 k  {  Tr \{ 
\gamma_5 \gamma^{-,-,+} (\st{k}+ \st{k}_1 + \st{q}_2 +\st{k}_3) 
\gamma_5 \gamma^{-,-,-} ~\st{k}~ 
\gamma_5 \gamma^{-,-,-}(\st{k}- \st{k}_2 + \st{q}_1 + \st{k}_3 ) \}
\over  (k + k_1 + q_2 + k_3 )^2  
~k^2 ~
 (k - k_2 + q_1 + k_3)^2 }  ~+ ~ \cdots }
\auto\label{580}
$$
where
$$
\gamma^{\pm,\pm,\pm} ~=~ \gamma^{\mu}\cdot n^{\pm,\pm,\pm}_{ \mu} ~,~~~~
n^{\pm,\pm,\pm \mu} ~= ~ (1,\pm1,\pm1,\pm1)
\auto\label{g64}
$$
The contributions to $R$ not shown explicitly in (\ref{580}) do not have
a $\gamma_5$ at all three vertices of the triangle diagram. The particular
$\gamma$-matrix projections appearing depend on the choice of transverse
co-ordinates. If the anomaly is present in $R$, however, we expect it to be
independent of this choice. We should
emphasize that while we have written (\ref{580}) as a function of
four-dimensional momenta, the $k_i$ are restricted to be 
two-dimensional spacelike momenta (plus longitudinal components
determined by the mass-shell conditions for the on-shell quarks) and 
the $q_i$ have the restricted form given by (\ref{np3}). These restrictions 
plays a crucial role in determining whether the anomaly can occur in a physical
region reggeon interaction.

A reggeon diagram amplitude 
that represents right-hand cuts in the unphysical triplet 
$\{s_{13'}, s_{32'}, s_{21'}\}$ and has two reggeons in each $t$-channel, 
each with trajectory $\alpha(t) = 1 +~O(g^2)$, has the form\cite{arw99}
$$
\eqalign{ &~~\prod_i\int { d^2k_i \over 
sin \pi \alpha (k_i^2)  sin \pi \alpha ((Q_i -k_i)^2)  } 
~~~~\beta(k_1,k_2,k_3,Q_1,Q_2,Q_3)\cr 
&\biggl[ ~(s_{13'})^{[\alpha (k_1^2)+\alpha ((Q_1 -k_1)^2) + 
\alpha (k_3^2)+\alpha ((Q_3 -k_3)^2) -
\alpha (k_2^2)-\alpha ((Q_2 -k_2)^2) -1]/2} \cr
& ~~~~~~~~ (s_{32'})^{[\alpha (k_3^2)+\alpha ((Q_3 -k_3)^2) + 
\alpha (k_2^2)+\alpha ((Q_2 -k_2)^2) -
\alpha (k_1^2)-\alpha ((Q_1 -k_1)^2) -1]/2} \cr
& ~~~~~~~~~~~~(s_{21'})^{[\alpha (k_1^2)+\alpha ((Q_1 -k_1)^2) + 
\alpha (k_2^2)+\alpha ((Q_2 -k_2)^2) -
\alpha (k_3^2)-\alpha ((Q_3 -k_3)^2) -1]/2} \biggr]~ ~~\bigg/ \cr
&~~~~\biggl[[sin {\pi \over 2} [\hbox{${\scriptstyle\alpha (k_1^2)
+\alpha ((Q_1 -k_1)^2) + 
\alpha (k_3^2)+\alpha ((Q_3 -k_3)^2) -
\alpha (k_2^2)-\alpha ((Q_2 -k_2)^2)}$}] \cr
&~~~~~~~~~~ sin {\pi \over 2} [\hbox{${\scriptstyle \alpha (k_3^2)
+\alpha ((Q_3 -k_3)^2) + 
\alpha (k_2^2)+\alpha ((Q_2 -k_2)^2) -
\alpha (k_1^2)-\alpha ((Q_1 -k_1)^2)}$} ] \cr
& ~~ ~~~~~~~~~~~~sin {\pi \over 2} [\hbox{${\scriptstyle \alpha (k_1^2)
+\alpha ((Q_1 -k_1)^2) + 
\alpha (k_2^2)+\alpha ((Q_2 -k_2)^2) -
\alpha (k_3^2)-\alpha ((Q_3 -k_3)^2)}$} ] \biggr]}
 \auto \label {2ra1}
$$
$$
\centerunder{$\sim$}{\raisebox{-6mm}{$ g^2 \to 0$}}~
(s_{13'})^{1/2}(s_{32'})^{1/2}(s_{21'})^{1/2}
~\prod_i\int{ d^2k_i \over 
k_i^2  (Q_i -k_i)^2  }
~~\beta_(k_1,k_2,k_3,Q_1,Q_2,Q_3) ~~
\auto\label{2ra}
$$
Taking the triple discontinuity in $s_{13'}$, $s_{32'}$
and $s_{23'}$ removes the poles due to the sine factors in the square 
bracket, but leaves the $g^2 \to 0$ limit unchanged. Since   
the triple discontinuity is unphysical and
of the kind discussed in the previous Section,
according to the discussion in
\cite{arw99}, the ``six-reggeon interaction vertex''
$\beta_(k_1,k_2,k_3,Q_1,Q_2,Q_3)$ could contain the anomaly.

Writing
$$
P_{1^+} P_{2^+} P_{3^+}~\equiv ~(s_{13'})^{1/2}(s_{32'})^{1/2}(s_{21'})^{1/2}
\auto\label{p+inv}
$$
and comparing with (\ref{2ra}) we see that (\ref{211}) 
could be identified as a lowest-order contribution to such a
reggeon diagram amplitude if the reduced feynman
diagram amplitude of Fig.~3.1(c) is identified as a reggeon vertex, i.e.
$$
R^6(Q_1,Q_2,Q_3, k_{1},Q_1- k_{1},\cdots)~\equiv ~
\beta(k_1,k_2,k_3,Q_1,Q_2,Q_3)
\auto\label{6rv}
$$
Therefore, if (\ref{211}) does represent a contribution to the
asymptotic behavior of the feynman diagram in Fig.~3.1(a) it could   
contribute to a triple-regge amplitude 
of the form of (\ref{2ra1}). Note that while the right-side of (\ref{p+inv})
clearly has a triple discontinuity in $\{s_{13'}, s_{32'}, s_{21'}\}$,
the left-side does not. The equivalence of the two sides is only determined
if higher-order terms in (\ref{2ra1}) appear and add to (\ref{211}) in 
the appropriate manner. Such terms are contributed by what we refer to as 
reggeization diagrams, both in the Introduction and in the following.   

As we have emphasized, the amplitude (\ref{580}) representing Fig.~3.1(c) 
is the full four-dimensional
triangle diagram amplitude except that special $\gamma$-matrices
appear at the vertices and
only combinations of (essentially) two-dimensional transverse  
momenta flow through the diagram.  As discussed in \cite{arw99}, the $\gamma$-matrix
couplings are appropriate to produce the anomaly but 
whether the necessary 
momentum configuration can occur within a physical region 
and provide a physical region inra-red divergence is a non-trivial
and subtle question that depends crucially on the choice of propagator
poles used to put lines on-shell, as we now describe.

The divergence of the (massless) triangle diagram occurs\cite{cg,arw99} 
when a single light-like momentum flows through the diagram and
all other momenta are spacelike and scaled to zero.
Such a momentum configuration for the reggeon interaction $R$ appears to be 
(essentially uniquely) realized by that of the 
full feynman diagram shown in Fig.~3.2(a).
\newline \parbox{4.3in}{
\begin{center}
\epsfxsize=4.1in
\epsffile{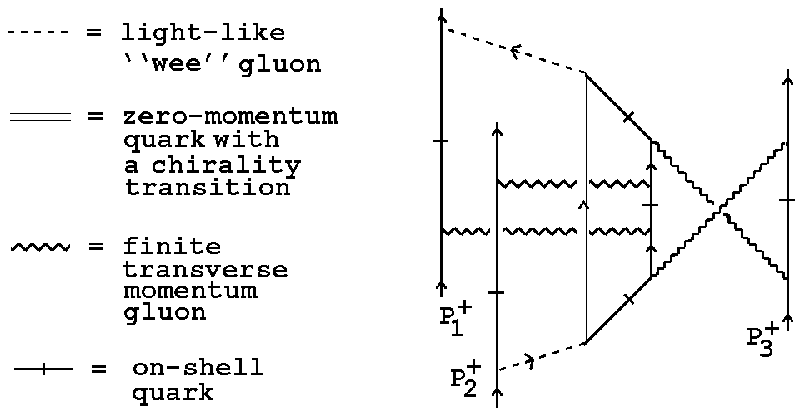}
\newline (a)
\end{center}}
\parbox{1.7in}{
\begin{center}
\leavevmode
\epsfxsize=1.5in
\epsffile{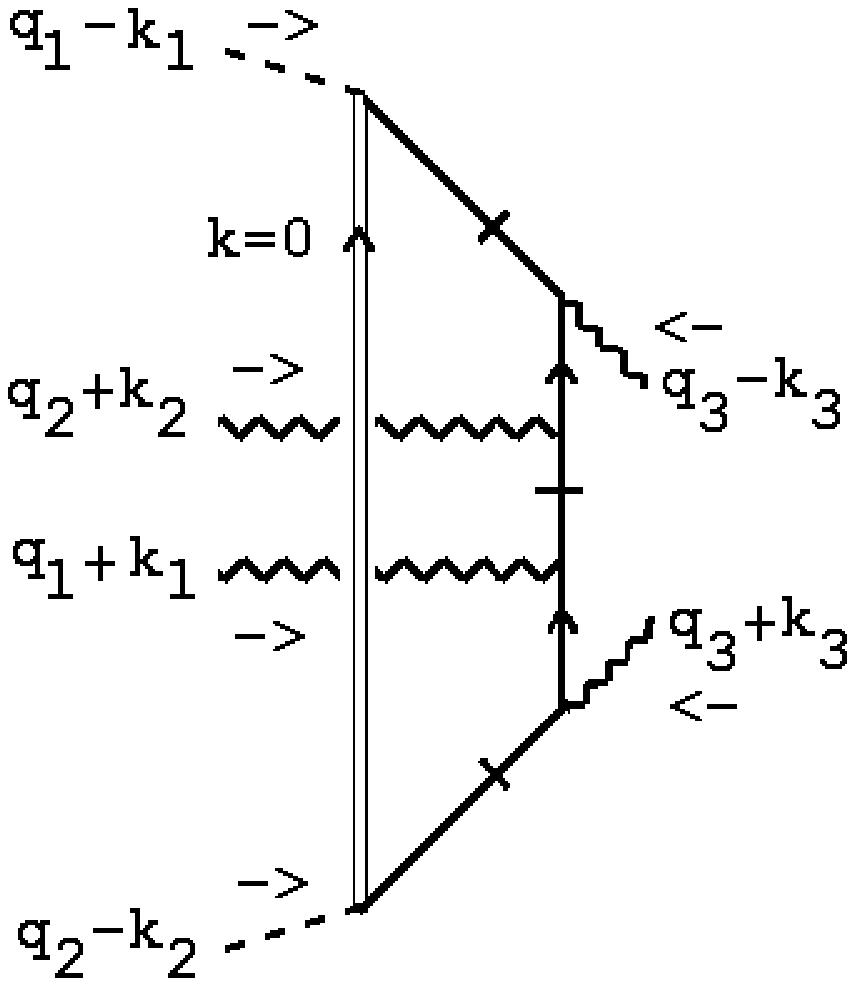}
\newline (b)
\end{center}}
\begin{center}
Fig.~3.2 The basic anomaly process.
\end{center}
If we label the momenta entering the reggeon interaction 
as in Fig.~3.2(b),
an explicit configuration for Fig.~3.2(a) is
$$
q_1-k_1~=~(2l,2l,0,0)~, ~~~
q_2-k_2~=~(-2l,0,-2l,0)
\auto\label{chm1}
$$ 
together with
$$
\hat{q}_1=- \hat{q}_2 =l ~~~~q_{13}=-q_{23} ~~~~ q_{12}=q_{21}=0
\auto\label{chm2}
$$
This determines $k_1$ and $k_2$ and also gives
$$
q_3~=~- (q_1+q_2)~= ~(0,-l,l,0)
\auto\label{chm3}
$$
If we then take
$$
k_3~=~l(0, 1 -2~ \cos{\theta}~, 1 - 2~\sin{\theta}~, 0) 
\auto\label{chm4}
$$
the light-cone momentum 
$$
-~2l(1, \cos{\theta},\sin{\theta}, 0)
\auto\label{chm30}
$$
flows along the two vertical non-hatched lines in Fig.~3.2(b). 
It is straightforward
to check that all three of the hatched lines are on mass-shell.
If spacelike momenta of $O(q)$ are added to the momentum 
configuration (\ref{chm1})-(\ref{chm30}) and the limit $q \to 0$ is taken
the presence of the anomaly divergence will lead to the behavior
$$
R^6(l,\theta,q) ~~\centerunder{$\sim$}{\raisebox{0.3mm}{$q \to 0$}}  
~~ {(1 - \cos{\theta} - \sin{\theta})^2
~l^2 \over q } 
\auto\label{anomd}
$$

Apart from the reversal of direction for $P_3$,  the process represented by 
Fig.~3.2(a) 
is what we called ``the basic anomaly process'' in \cite{arw99}.
The scattering should be thought of as taking place with the time axis
vertical on the page. The space axes are not, of course, represented 
accurately since each of the external quarks is traveling along orthogonal
directions. The dashed lines indicate light-like (``wee
parton'') gluons with finite momenta parallel to the 
incoming/outgoing quark that they are emitted/absorbed by.
A zero-momentum quark (indicated by the open line) 
is emitted by the incoming wee-parton gluon and  
is absorbed by the outgoing wee-parton gluon. 
The first and last gluon 
interactions of the antiquark rotate it's incoming/outgoing
lightlike momentum to/from the
triangle light-like momentum associated with the infra-red divergence.
The two intermediate gluons carry equal but opposite transverse 
momenta. Their combination provides a 
forward scattering of the antiquark that, most importantly,
includes a spin flip (the momentum factor for which reduces what would be a $1/q^2$
factor in (\ref{anomd}) to $1/q$). 

The zero momentum quark is produced by one wee
gluon and absorbed by the other, allowing the chirality
transition produced by the anomaly
to compensate for the spin flip of the antiquark.  
Note, however, that when the wee gluons are massless,
the scattering processs represented by Fig.~3.2 is physical only when the 
quark and antiquark involved are also massless. In addition, as we 
noted in the Introduction, the 
anomaly infra-red divergence involves both poles of 
the zero momentum quark propagator. According to the helicity analysis
of \cite{cg} the vertices coupling to the propagator 
should be symmetrically interpreted
as describing either the simultaneous production of the two states 
in the propagator or their simultaneous 
absorption. Therefore, if (the infra-red divergence analysis that we 
ultimately employ to define physical states and amplitudes should
require that) we interpret 
the massless scattering as entering the physical region
with the time ordering implied by Fig.~3.2, we are implicitly
assuming the presence of a non-perturbative background gauge field. The
background field would be needed to
produce the necessary spectral flow at one vertex 
that is required to interpret the process as a chirality transition. 
 
While the mass-shell
conditions are indeed satisfied by (\ref{chm1})-(\ref{chm30}), we must now discuss
the important subtlety as to which propagator pole is chosen. 
With the momenta given by  (\ref{chm1})-(\ref{chm30}),
the energy component of each of the three hatched lines in Fig.~3.2(b)
has the same sign.
Since the exchanged gluons carry only spacelike momenta, it is clear that 
this must be the case. Equivalently, the
on-shell states in the loop must be treated symmetrically in
that, if the zero momentum state is an antiquark (quark), 
all hatched lines must be quarks (antiquarks).
We refer to this as  the ``all quarks requirement''. 
As we already remarked on in the Introduction, and as is discussed at length 
in \cite{arw99}, the only practicable calculational method
to determine whether
a given combination of on-shell lines contributes to the triple-regge behavior
(after all diagrams are added) is the dispersion relation method that 
we outline very briefly below. In this 
approach all on-shell lines 
in a reggeon interaction result directly from the taking of a 
triple asymptotic 
discontinuity. ``Real part'' interactions with the 
same on-shell lines may be
generated when the full dispersion relation is written
or, equivalently, multi-regge theory is used\cite{arw99} 
to convert the triple discontinuity to a full amplitude. 

In fact, the ``all quarks requirement''  
is (as we shall see from the examples we
discuss below) very difficult to obtain in a reggeon interaction derived 
from a multiple discontinuity. We believe, however, that it is an essential
requirement for a reggeon interaction to contain a physical region divergence
produced by the anomaly, i.e. some variant of the ``basic anomaly process''
must be involved. We recognized in \cite{arw99} that 
the necessary triple discontinuity is not present in the diagram of Fig.~3.1
but we suggested that nevertheless it may be present in the higher-order 
reggeization
diagrams that produce the reggeization of the gluons and so the basic anomaly 
process of Fig.~3.2 may be required as a real part interaction
(via the equivalence (\ref{p+inv})).
In fact, we will show in the remaining part of
this paper that this is not the case. Instead,
to satisfy the all quarks requirement,
there must be at least two wee gluons (instead of one) either emitted, or 
absorbed, in the basic anomaly process. Ultimately this implies that reggeon
interactions with the quantum numbers of the winding number current must
be involved. 
As we emphasized in the Introduction (and also discussed in \cite{arw99}) 
we do not expect the anomaly divergence to be present in the scattering of
elementary quarks and/or gluons after all diagrams are summed. Rather,
as we briefly comment on in Section 6, we 
expect it to be present when  
the basic process is generalized to describe the scattering of 
the particular multi-regge states that ultimately form
bound states. The corresponding $G_i$ will then appear in 
a generalization of (\ref{211}) and the 
wee partons involved will be a crucial characteristic of  
scattering states. Also the chirality transitions produced (and the implicit 
spectral flow) will be an essential part of scattering processes.

In general, an asymptotic dispersion relation\cite{arw00} gives the leading 
multi-regge behavior of an amplitude as
a sum over multiple discontinuity contributions 
allowed by the Steinmann relations. For the particular case (described in
detail in \cite{arw99}) of 
the triple-regge 
behavior of a six-point amplitude we can write 
$$
M_6(P_1,P_2,P_3,Q_1,Q_2,Q_3)~ =~ 
\sum_{\cal C} M_6^{\cal C}(P_1,P_2,P_3,Q_1,Q_2,Q_3)
~+~M_6^0~,\auto\label{dis}
$$
where $M_6^0$ contains all non-leading 
triple-regge behavior, double-regge behavior, etc. and the sum is
over all triplets ${\cal C}$ of 
asymptotic 
cuts in non-overlapping (large) invariants. For each triplet ${\cal C}$, 
say ${\cal C}= (s_1,s_2,s_3)$, we can write 
$$  
\eqalign{M_6^{\cal C}(P_1,P_2,P_3,Q_1,Q_2,Q_3)~=~{1\over (2\pi i)^{3}}  &~~\int
ds'_1 ds'_2 ds'_{3} ~~{\Delta^{\cal C} \over
(s'_1-s_1)(s'_2-s_2)(s'_{3}-s_{3})} }
\auto\label{dis2}
$$
where $\Delta^{\cal C}$ is the triple discontinuity.

The triple discontinuities are of three
kinds corresponding to the three tree diagrams of Fig.~3.3. 
There are 24 corresponding to Fig.~3.3(a),
12 corresponding to Fig.~3.3(b), 
and 12 of the  Fig.~3.3(c) kind - including those described by Fig.~2.3. 
Those of Fig.~3.3(a) and (b), occur in the physical
regions, while those corresponding to Fig.~3.3(c) are all unphysical 
triple discontinuities of the kind discussed in the last Section.
\begin{center}
\leavevmode
\epsfxsize=2.2in
\epsffile{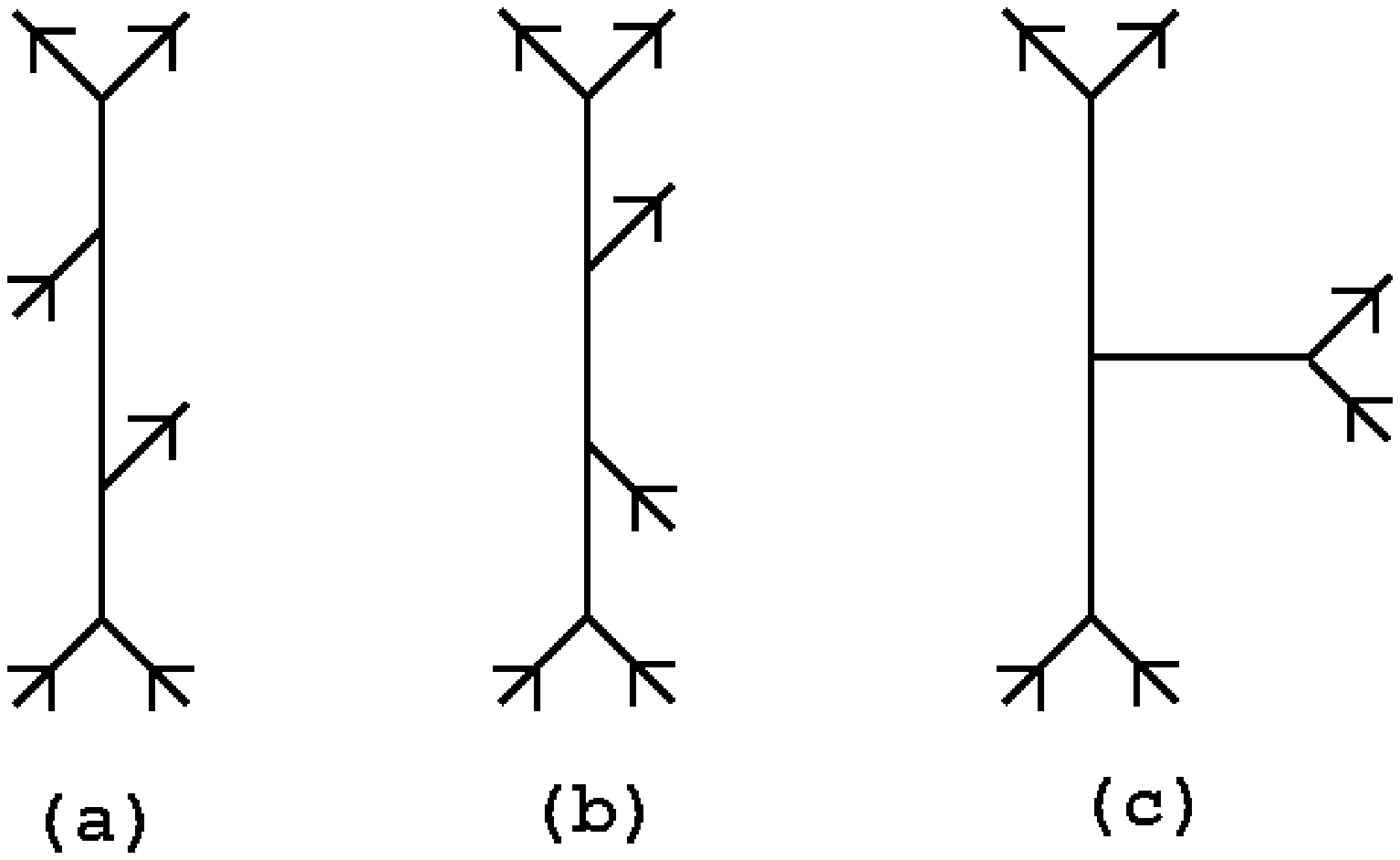}

Fig.~3.3 Tree Diagrams for triple discontinuities.
\end{center}

As we discussed in \cite{arw99}, the diagram of Fig.~3.1(a) has physical region
triple discontinuities of both the Fig.~3.3(a) and (b) kinds, although 
neither gives leading triple-regge behavior. Unphysical discontinuities
are more complicated to discuss. If the usual cutting rules hold,
the diagram of Fig.~3.1(a) has no asymptotic triple 
discontinuities corresponding to Fig.~3.3(c), but rather has
only double discontinuities. To see this, 
consider cutting the diagram as in Fig.~3.4, superficially giving an
$\{s_{13'}, s_{32'}, s_{21'} \}$  triple discontinuity.
\begin{center}
\epsfxsize=2in
\epsffile{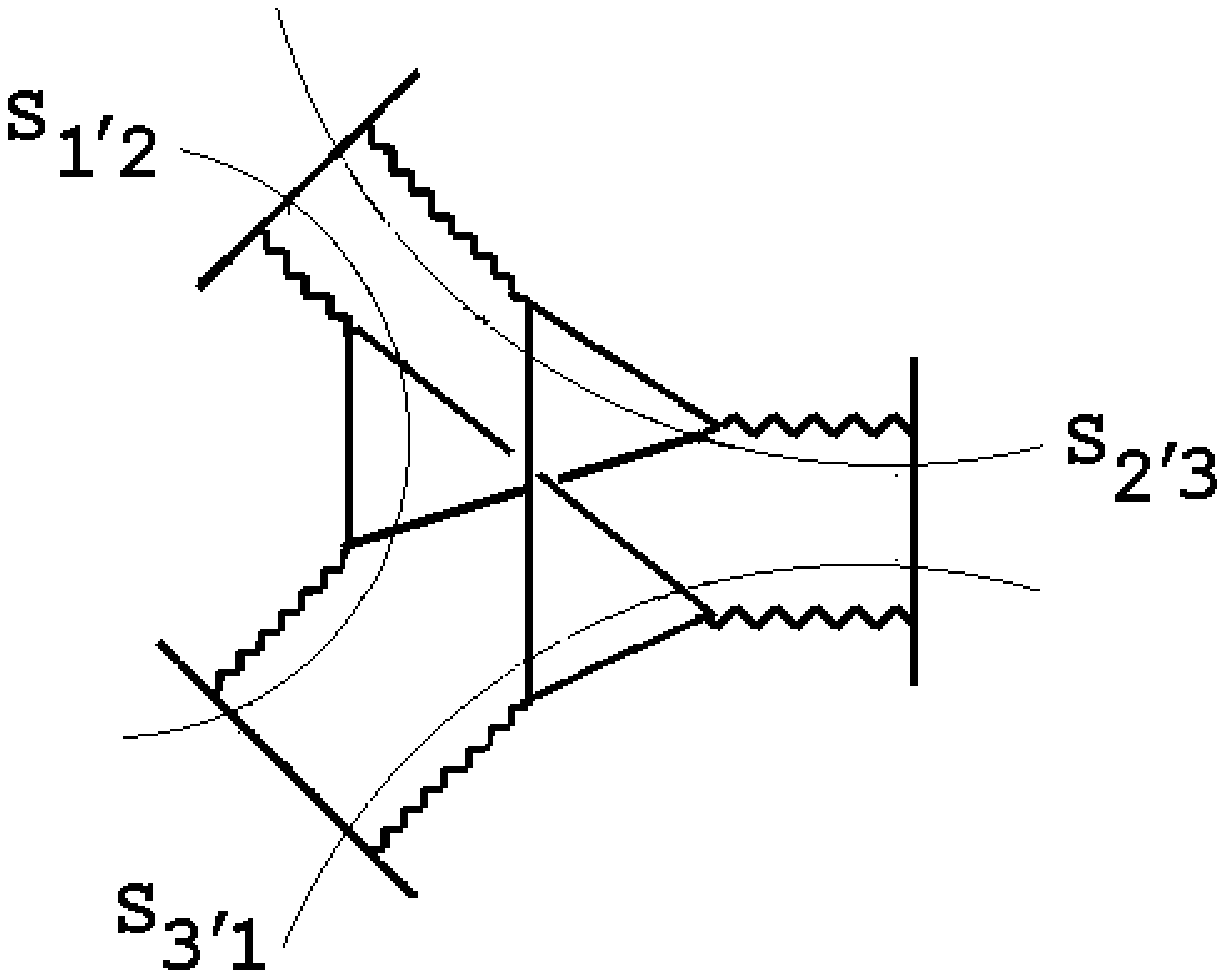}

Fig.~3.4 An unphysical triple discontinuity?
\end{center}
In fact, 
taking any double discontinuity clearly cuts all the available lines, implying 
that there is no independent third discontinuity that can be taken.

It is not clear a-priori 
that the cutting rules do apply to unphysical discontinuities. However, we will
show directly in Section 5 that there is no symmetric
triple discontinuity present (giving the desired common energy 
component sign) in the diagram of Fig.~3.1. 
Therefore, as we described above, whether there is an anomaly contribution
from diagrams of this kind depends on whether the necessary triple
discontinuities are present when  
reggeization effects appear.
In \cite{arw99} we noted only that such discontinuities 
appeared to be present in reggeization diagrams 
but did not discuss the structure of such diagrams in any detail. 

As an example of a diagram that should produce reggeization, 
consider that shown in Fig.~3.5  
\begin{center}
\epsfxsize=4.5in
\epsffile{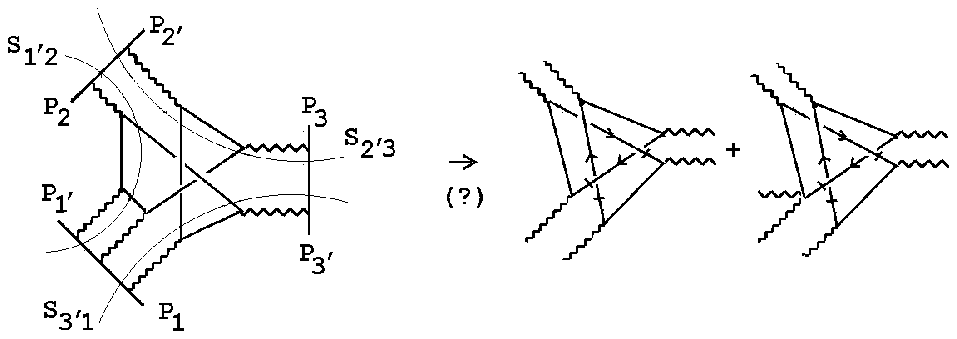}

Fig.~3.5 A diagram with an unphysical triple discontinuity.
\end{center} 
in which one of the gluons in the diagram of Fig.~3.4 is replaced by 
two-gluon exchange - potentially  
giving the one-loop contribution to the trajectory
function of the original gluon. The thin lines again indicate how an
unphysical $\{s_{13'}, s_{32'}, s_{21'} \}$ discontinuity would be taken.
In such a contribution  
the corresponding six reggeon interaction, together with a remnant 
seven reggeon interaction, would be generated by putting the cut
lines on-shell. The discontinuity is clearly not symmetric and, in addition,
if the particles put on-shell by a 
discontinuity must be either all ``incoming'' or all ``outgoing'' 
(this is the ``$+\alpha$'' condition that is part of  
the normal cutting rules) then the energy components of
the on-shell lines in the quark loop (apart from that potentially giving
the reggeization contribution) can not have the common sign 
required for the anomaly. This is because 
both the $s_{2'3}$ and $s_{3'1}$ cuts involve two 
of the relevant quark loop lines which must, therefore, be either incoming or
outgoing quark/antiquark pairs. This requirement then 
eliminates the possibility that such lines are all antiquarks, or all quarks.

In the next Section we will confirm by direct calculation that the diagram of
Fig.~3.5 does not have the triple discontinuity needed to give the anomaly.
Consequently, the reggeon interactions generated do not contain the anomaly.
Although a more complete analysis of all diagrams should be given, this 
essentially determines that the anomaly process of Fig.~3.2 is not 
generated as a ``real part interaction'' when higher-order 
reggeization effects are included. 

To obtain a symmetric triple discontinuity 
in which the normal cutting rules could potentially give the anomaly amplitude 
associated with  Fig.~3.2 , we consider the
high-order diagram shown in Fig.~3.6(a)
\newline \parbox{1.7in}{
\begin{center}
$~$
\newline $~$
\newline $~$
\newline $~$
\newline \epsfxsize=1.2in
\epsffile{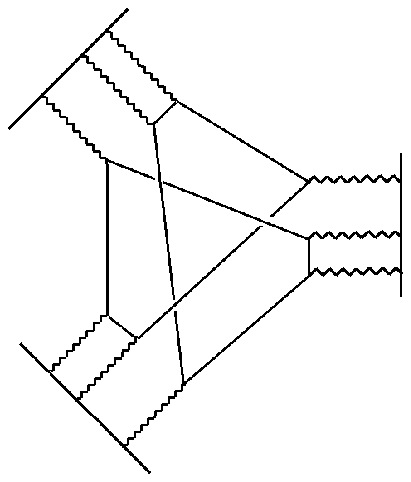}
\newline $~$
\newline $~$
\newline(a)
\end{center}}
\parbox{4.3in}{
\begin{center}
\epsfxsize=3.8in
\epsffile{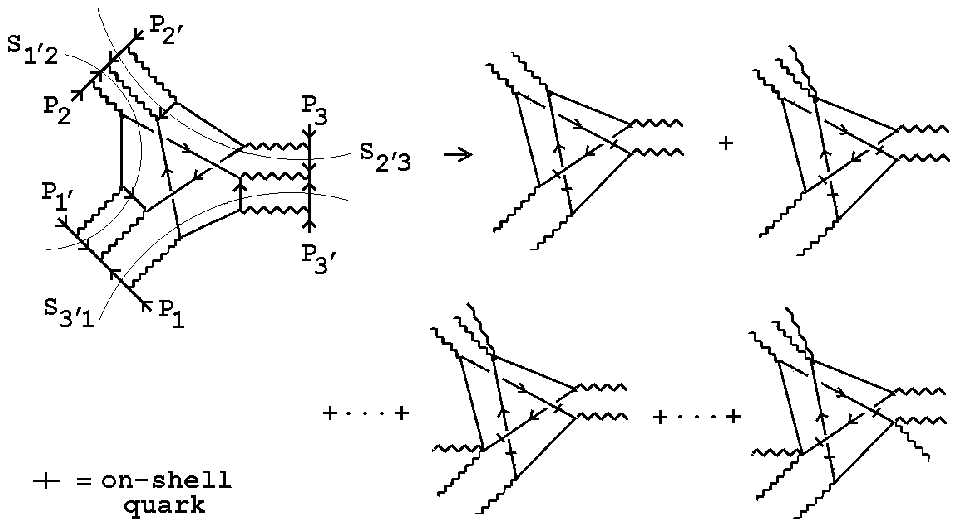}
\newline $~$
\newline(b)
\end{center}}
\begin{center}
Fig.~3.6 (a) A diagram with a symmetric 
unphysical triple discontinuity
\newline (b) expected reggeon interactions.
\end{center}
in which there are three gluons in each
$t$-channel. A triple discontinuity in $\{s_{1'2}, s_{2'3}, s_{3'1} \}$ 
is obtained by cutting the diagram as indicated in Fig.~3.6(b). 
The closed loops involving two-gluon exchange could give
both one loop contributions to the corresponding one reggeon trajectory
function and the leading contribution of a two reggeon state. A-priori,
therefore, we expect the diagram to contribute to the six-, seven-,
eight- and nine-reggeon interaction as illustrated.
Since the triple discontinuity is manifestly symmetric we again might 
expect the anomaly to appear in the six-reggeon interaction, just
as anticipated by the lower-order amplitude of Fig.~3.1(a).

For consistency with our previous discussion,
the anomaly should not (and does not) appear quite so simply.
As will be clear after we carry out the explicit 
evaluation of asymptotic discontinuities in Section 5,  
the triple discontinuity of Fig.~3.6(b) requires a 
particular routing of the internal loop momenta. For this routing 
the numerators of the cut quark propagators do not give the combination of
$\gamma_5$ interactions needed for the anomaly. 
In fact, the anomaly does occur within 
a reggeon interaction generated by the diagram of Fig.~3.6(a) but  
only when the unphysical discontinuities are actually taken 
as shown in Fig.~3.7. It then occurs in the
nine-reggeon interaction obtained, as illustrated, by putting lines on-shell.
\begin{center}
\epsfxsize=3.2in
\epsffile{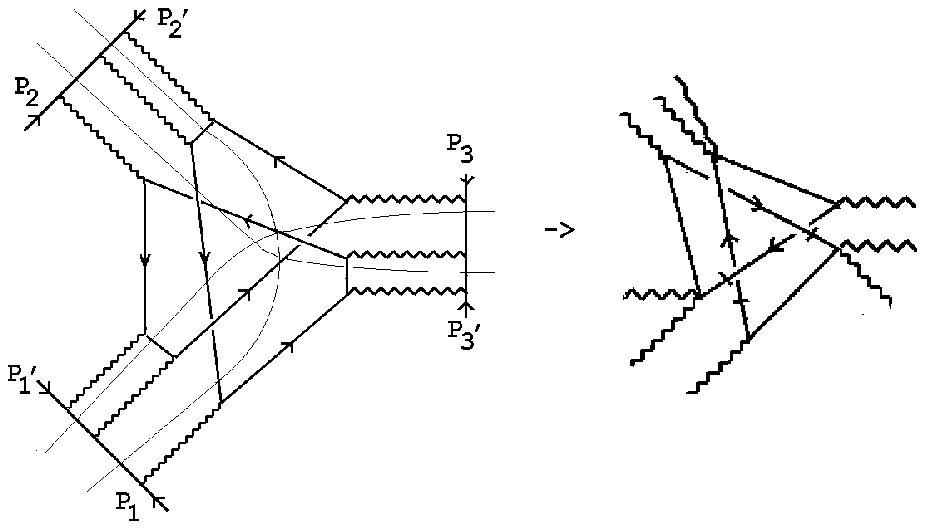}

Fig.~3.7 Another cutting of Fig.~3.6(a).
\end{center}
Note that the discontinuity lines in Fig.~3.7 cross each other. This is 
possible because the particles contributing to each discontinuity do not
all have the same time direction. 
To evaluate a multiple discontinuity of this kind
we must develop direct methods to
compute asymptotic discontinuities.

In the reggeon interaction of Fig.~3.7  there are 
three reggeons in each $t$-channel and each reggeon state is 
``vector-like'' in that it has (close to) unit angular momentum 
and appears in odd-signature amplitudes.
As discussed in \cite{arw99},
to avoid cancelation of the anomaly by transverse momentum integrations
each reggeon state should have abnormal parity. 
Therefore, according to the above discussion, the simplest
interaction in which 
the anomaly could appear is the nine-reggeon interaction in which 
each reggeon state
is vectorlike, composed of (at least) three gluons, and has abnormal parity.
If, in addition, each reggeon state has zero color 
then all three states carry anomaly current quantum numbers.
It is surely remarkable that we are led directly to 
the anomaly current by looking for the anomaly within   
reggeon interactions.

In fact, the analyticity properties of amplitudes require \cite{arw99} 
that the anomaly appears only when 
signature conservation is also satisfied, which it is not
not if all three reggeon states carry odd signature. Therefore, to avoid
cancelation when further diagrams are added, an additional (reggeized) gluon
must be present in one $t$-channel. This gives 
only a relatively trivial modification of Fig.~3.7 and the 
analysis that follows. In this paper, we are not interested
in determining when the anomaly ultimately
survives after all diagrams 
are summed. We are satisfied just to find diagrams in which our asymptotic 
discontinuity analysis determines that the anomaly is definitively present
in the extracted reggeon interaction.
This already requires that we go to the complexity of Fig.~3.7.

\newpage

\mainhead{4. LIGHT-CONE ANALYSIS OF ASYMPTOTIC DISCONTINUITIES}

In the next Section we will analyse triple-regge asymptotic discontinuities
and will use a generalization of the simple light-cone analysis
that we develop in this Section.
Consider the box-diagram illustrated in Fig.~4.1. 
\begin{center} 
\leavevmode
\epsfxsize=2.5in
\epsffile{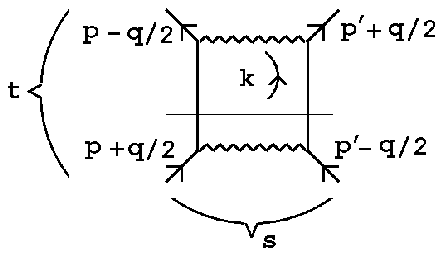}

Fig.~4.1 The box diagram.
\end{center}
Initially we ignore the role played by 
numerators and so we consider, in the notation shown,
$$
\eqalign{             
I(s,t,m^2) = &\int d^4k \left[k^2-m^2+i\epsilon\right]^{-1}
\left[\left(p-{q\over 2}+k\right)^2-m^2+i\epsilon\right]^{-1}\cr
&\times \left[(q-k)^2-m^2+i\epsilon\right]^{-1}\left[\left(
p' +{q\over 2}-k\right)^2-m^2+i\epsilon\right]^{-1}.}
\auto\label{lcan1}
$$
This integral is, of course, a function of invariants only even though it is
specified using four momenta. Indeed, we can evaluate the integral using
complex, unphysical, momenta that give physical values of the invariants,
provided we are careful to define the integral via analytic continuation from
the appropriate physical momentum region. Our purpose in this Section is to
discuss momentum dependence of this kind for the simplifying
case of the leading asymptotic behavior, in a manner that we can apply 
to much more complicated diagrams in Section 5.

For illustrative purposes we set both $q=0$ and $m=0$ in (\ref{lcan1}) 
and ignore infra-red divergences. We can then write
$$
I(s) ~= ~\int d^4k \left[k^2+i\epsilon\right]^{-2}
\left[\left(p +k\right)^2 +i\epsilon\right]^{-1}
\left[\left(p'  -k\right)^2+i\epsilon\right]^{-1}
\auto\label{lcan2}
$$
We choose a particular Lorentz frame and introduce light-cone co-ordinates 
such that 
$$
\eqalign{
p&~= ~\left({P_+ \over 2}~,{P_+ \over 2}~,~\til{0}\right) 
~ +O\left({1\over  s}\right), ~~~~~P_+ ~\sim ~ s ~\to ~ \infty \cr
p'&~=~  \left({P_+'+ P_-' \over 2}~, {P_+'- P_-' \over 2}~,~
\underline{p}_{\perp}'~\right)   }
\auto\label{lcan3}
$$
so that $s~ = P_+P_-'~[1 + O(1/ s)]~$. We can then write
$$
\eqalign{I(s)~\centerunder{$\large\sim$} {\raisebox{-3mm} 
{$\scriptstyle s\to \infty$}}~~
{1\over 2}\int &                                        
d^2\underline{k}_{\,\perp} dk_+dk_- \left[ k_+k_- -k^2_{\,\perp} + 
i\epsilon\right]^{-2}~
\bigg[ \left( k_+ + P_+ \,\right) k_-
- \underline{k}^2_{\,\perp}
+i\epsilon\bigg]^{-1}  \cr
& \times~
\left[\left(k_+ -P_+' \right)\left(k_- -P_-' \right)- 
(\underline{k}_{\,\perp} - \underline{p}_{\,\perp}')^2
+i\epsilon\right]^{-1} }
\auto\label{lcan4}
$$

To obtain a non-zero answer by closing the $k_+$ contour, with $k_-$ and 
$k_{\perp}$ fixed, the three poles given by the three square brackets of 
(\ref{lcan4}) must
not be on the same side of the contour. This requires $ 0 <k_-< P_-'$ and,
in this case, the $k_+$ contour can be closed to pick up only
the pole in the last bracket. This gives 
$$
k_+ ~=~ P_+' +
{(\underline{k}_{\,\perp} - \underline{p}_{\,\perp}')^2
- i\epsilon \over \left(k_- -P_-' \right)}
\auto\label{lcan40}
$$
which is finite and so can be neglected compared to $P_+$. Note also that 
$$
k_- \sim 0,~~ {k_{\perp}}^2 \sim 0 ~~=> ~~k_+~\sim ~2k_0~ \sim 
~{{p'}^2 \over P_-'}
\auto\label{5an}
$$
(we will need this approximation for the analysis of 
Section 5). We thus obtain, 
$$
I(s)~\centerunder{$\large\sim$} {\raisebox{-3mm} 
{$\scriptstyle s\to \infty$}}~~\pi i\int 
d^2\underline{k}_{\perp} \left[ -k^2_{\perp} + 
i\epsilon \right]^{-2} ~\int_0^{P_-'} dk_-  
\left[k_- -P_-' \right]^{-1}\left[ P_+ k_-
- \underline{k}^2_{\perp}
+i\epsilon \right]^{-1} 
\auto\label{lcan5}
$$

We are specifically interested in the leading real and imaginary parts of 
(\ref{lcan5}). They are given by the logarithm generated by 
the pole factor containing $P_+$ as it approaches the $k_- = 0$ end-point of 
the integration. If we keep only the integration over $0 < k_- < \lambda P_-'$ 
and take $\lambda << 1$ so that we can 
make the approximation $k_- / P_-'~ \sim 0$ we obtain
$$
\eqalign{ I(s)~& \centerunder{$\large\sim$} {\raisebox{-3mm} 
{$\scriptstyle s\to \infty$}}~~
\pi i\int 
d^2\underline{k}_{\perp} \left[ -\underline{k}^2_{\perp} + 
i\epsilon \right]^{-2} ~
{1 \over P_-'} ~\int_0^{\lambda P_-'} dk_-  
\left( P_+ k_-
- \underline{k}^2_{\perp}
+i\epsilon \right)^{-1} \cr 
&~\sim ~{1 \over P_+ P_-'} ~[\log{(P_+P_-'\lambda  
 -\underline{k}^2_{\perp} + i\epsilon]~J_1(0)} \cr 
&~\sim ~{1 \over s} ~[\log{(s\lambda + i\epsilon]~J_1(0)}
~ \sim ~{1 \over s} 
~[\log{s} + i\pi]~J_1(0) }
\auto\label{lcan60}
$$
where $J_1(0) ~\sim ~ \int 
d^2\underline{k}_{\perp} \left[ -\underline{k}^2_{\perp} + 
i\epsilon \right]^{-2}$ is infinite, but would be finite if we added a 
mass to the particle propagators. 

As we have indicated, the sign of the imaginary part in (\ref{lcan60}) arises 
directly from the $i\epsilon$ prescription. To obtain the leading imaginary 
part or, equivalently, the leading behavior of the discontinuity in $s$,
it suffices to keep the $i\epsilon$ dependence while dropping the 
$ -\underline{k}^2_{\perp}$ dependence in the $k_-$ integral. 
(\ref{lcan60}) is, of course, independent of $\lambda$. It will, however, be 
useful to note the role of $\lambda$ with respect to the 
analytic structure of $I(s)$ in the $s$-plane. As illustrated in Fig.~4.2,
the finite end of the branch-cut asociated with the logarithm in 
(\ref{lcan60}) moves out as $\lambda \to 0$.
\begin{center}
\leavevmode
\epsfxsize=1.6in
\epsffile{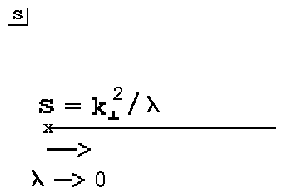}

Fig.~4.2 $\lambda$-dependence of the branch cut.

\end{center}
This is irrelevant to the asymptotic behavior and the 
``asymptotic discontinuity'' clearly remains unchanged. We will, nevertheless, 
be able to exploit this 
simple feature in evaluating multiple discontinuities in the next Section.
Also, although (\ref{lcan60}) is an 
invariant result, for
our purposes it will be useful to keep the dependence on both $P_+$ and $P_-'$
and discuss the dependence of the phase on $P_+$.

The initial $k_-$ integration contour for (\ref{lcan60})
is as shown in Fig.~4.3(a) with the pole at 
$ k_- = \underline{k}_{\perp}^2/ P_+$ indicated by a dot.
\begin{center}
\leavevmode
\epsfxsize=4.5in
\epsffile{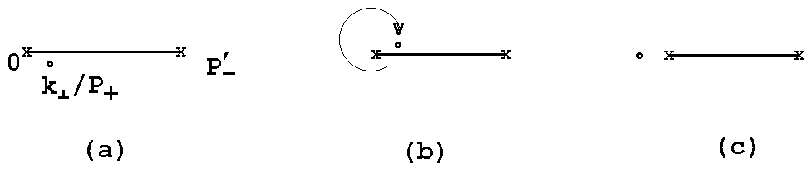}

Fig.~4.3 Integration Contours for (a) (\ref{lcan5})~ (b) 
~$P_+ \to~ e^{2\pi i}P_+$ ~(c) Fig.~4.4.
\end{center}
As $P_+$ (and therefore $s$)
completes a circle in the complex plane the pole 
moves around the end-point as illustrated in Fig.~4.3(b). The result is 
that the phase of the logarithm in (\ref{lcan60}) changes from $\pi$ 
to $-\pi$ and there is a net discontinuity of $2\pi i / s$, as is given 
directly by (\ref{lcan60}).
This is also the result that would be 
obtained by applying directly the standard
cutting rules to Fig.~4.1, cut by the thin line, if 
the $k_+$ and $k_-$ integrations are used to put the vertical lines 
on shell. The above discussion is simply an asymptotic analysis of 
how the two cut propagators pinch the integration region to generate 
a branch-point in $s$. Introducing $\lambda$ limits the integration 
region for the original integral such that the pinching only takes place for
$s \sim P_+ ~> \lambda$. Note also that the residue function $J_1(0)$,
multipying the logarithm in (\ref{lcan60}),
is directly obtained from the original box
diagram by putting the cut lines giving the discontinuity on-shell using 
the longitudinal momentum integrations. This is a very simple example 
(the simplest) of the relationship between a discontinuity and asymptotic
behavior.  

In evaluating unphysical (multiple) discontinuities in 
Section 5 we will not assume 
that the standard cutting rules apply. Instead we will directly analyse 
the discontinuities produced by logarithms. To understand how a discontinuity
generated by a logarithm can provide leading asymptotic behavior 
we note that the twisted diagram of Fig.~4.4, for
$q=0$, differs from that of Fig.~4.1 only by $P_+ \to -P_+$.
\begin{center}
\leavevmode
\epsfxsize=2in
\epsffile{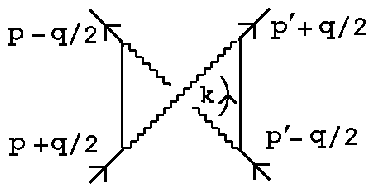}

Fig.~4.4 The twisted box diagram.
\end{center}
As a result,
the integration contour and pole position of Fig.~4.3(a) is replaced by 
that of Fig.~4.3(c). In this case a discontinuity is generated 
for $s< 0$. For
$s>0$ there is no phase generated by Fig.~4.4 and only the real
logarithms cancel when this diagram is added to that of Fig.~4.1. 
The leading behavior 
of the discontinuity in $s$, i.e the imaginary part, produced by the
diagram of Fig.~4.1 remains. 
This cancelation of the logarithms is very well-known, of course. 
It is also well known
that the cancelation fails when a non-abelian symmetry group is present
and that a consequence is the reggeization of the gluon.

We can briefly summarize the effect of adding numerators to (\ref{lcan1})
as follows.
First we note that the numerator of the internal fermion propagator 
carrying $P_+$ gives an additional $P_+$ factor of the form $\gamma_- P_+$. 
As a consequence, in (\ref{lcan60}), there is the replacement 
$$
~\int_0 dk_- ~
\left( P_+ k_- + \cdots \right)^{-1} ~~ \to ~\gamma_- P_+ ~ 
\int_0 dk_-  \left( P_+ k_- + \cdots \right)^{-1} ~\sim~ \log{P_+}
\auto\label{lcan61}
$$
and there is no inverse power of $P_+$. Also, each
coupling to a gluon gives a $\gamma$ matrix factor and since the external 
fermion lines are on-shell we can use the asymptotic form of 
the Dirac equation (i.e. $ \gamma_- P_+ \psi ~\sim m ~\psi $) to write
$$
\eqalign{ <P_+|\gamma_{\mu}\gamma_- \gamma_{\nu}|P_+>~& \sim ~
<P_+|{\gamma_- P_+ \over m} ~\gamma_{\mu}\gamma_- \gamma_{\nu}~
{\gamma_- P_+ \over m}|P_+> \cr
&  =~ <P_+|P_+ \gamma_- P_+ |P_+> / m^2~~
\sim P_+ ~/m }
\auto\label{coup}
$$
This gives another power of $P_+$ ($\sim s$) provided that the corresponding
factor of $P_-'$ is present in the finite momentum part of the scattering 
process. Not surprisingly this factor emerges from that part 
which would dominate if $P_-'$ were large. However, we want to 
emphasize that this selection is made only by the need to form a Lorentz 
invariant amplitude from the non-invariant large momentum process.

Finally we note that the above analysis goes through with very little
modification if we take both $m^2$ and $q$ to
be non-zero so that (\ref{lcan2}) will not be infra-red divergent. 

\newpage

\mainhead{5. UNPHYSICAL TRIPLE DISCONTINUITIES AND HIGHER-ORDER GRAPHS}

In this Section we generalize the analysis of the last Section to asymptotic 
triple discontinuities. The essential idea is that there is a well-defined 
leading-log result for each triple discontinuity, just as there was for
the single discontinuity in $s$ in the last Section, and that this can be found
from the leading-log calculation of an amplitude by keeping the $i \epsilon$
dependence of all logarithms. 

We begin by considering again the maximally non-planar graph
shown in Fig.~3.1. To understand 
how asymptotic discontinuities of the kind we are interested in 
arise, we first consider
a physical region discontinuity. To this end we interchange 
$P_1$ and $P_{1'}$ in (\ref{np3}) so that $P_{1'}$ and $P_{2}$ are the
momenta of incoming particles. For simplicity, we also set 
$Q_i=0,~ i=1,2,3$. This could cause confusion as to which invariants
discontinuities actually occur in. However, for the discontinuities
that interest us, we will be able to
avoid this issue. (As in the previous Section, adding both transverse 
momenta and masses to our discussion would not change the essential features 
of the analysis, but would eliminate gluon 
infra-red divergences. We will discuss,
at some points, the general effect of adding transverse momenta.)
Therefore we write, asymptotically,  
$$
\eqalign{ P_{1'}~\to &~- P_{1}~= ~(p_{1'},p_{1'},0,0)~,~~p_{1'} \to \infty \cr
P_2~\to &~- P_{2'}~= ~(p_2,0,p_2,0)~,~~p_2 \to \infty \cr
P_3~\to &~ -P_{3'}~= ~(p_3,0,0,p_3)~,~~p_3 \to \infty  }
\auto\label{pas}
$$

Since we will ultimately be looking for a symmetric triple discontinuity,
we consider only routes for the 
internal loop momenta of Fig.~3.1 that are completely symmetric with 
respect to the three external loops. There is essentially only one possibility.
The two apparently distinct possibilities   
illustrated in Fig.~5.1 are related by interchanging the primed and 
unprimed external momenta.
We will also want to make a symmetric choice for 
the quark lines we place on shell. Although we will not discuss the anomaly 
in detail until the next Section, 
we note that a product of three orthogonal $\gamma$-matrices
must be associated with the process of putting on-shell 
each internal quark line.
To achieve this it is necessary to put on-shell, symmetrically,  
the internal lines in Fig.~5.1(a) 
along which a single loop momentum flows. Therefore, we 
consider only such lines in the following.
\newline \parbox{3in}{ 
\begin{center}
\epsfxsize=2.2in
\epsffile{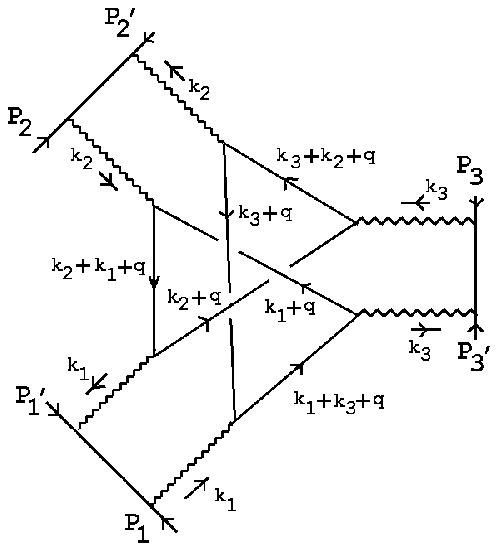}
\newline (a)
\end{center}}
\parbox{3in}{ 
\begin{center}
\epsfxsize=2.2in
\epsffile{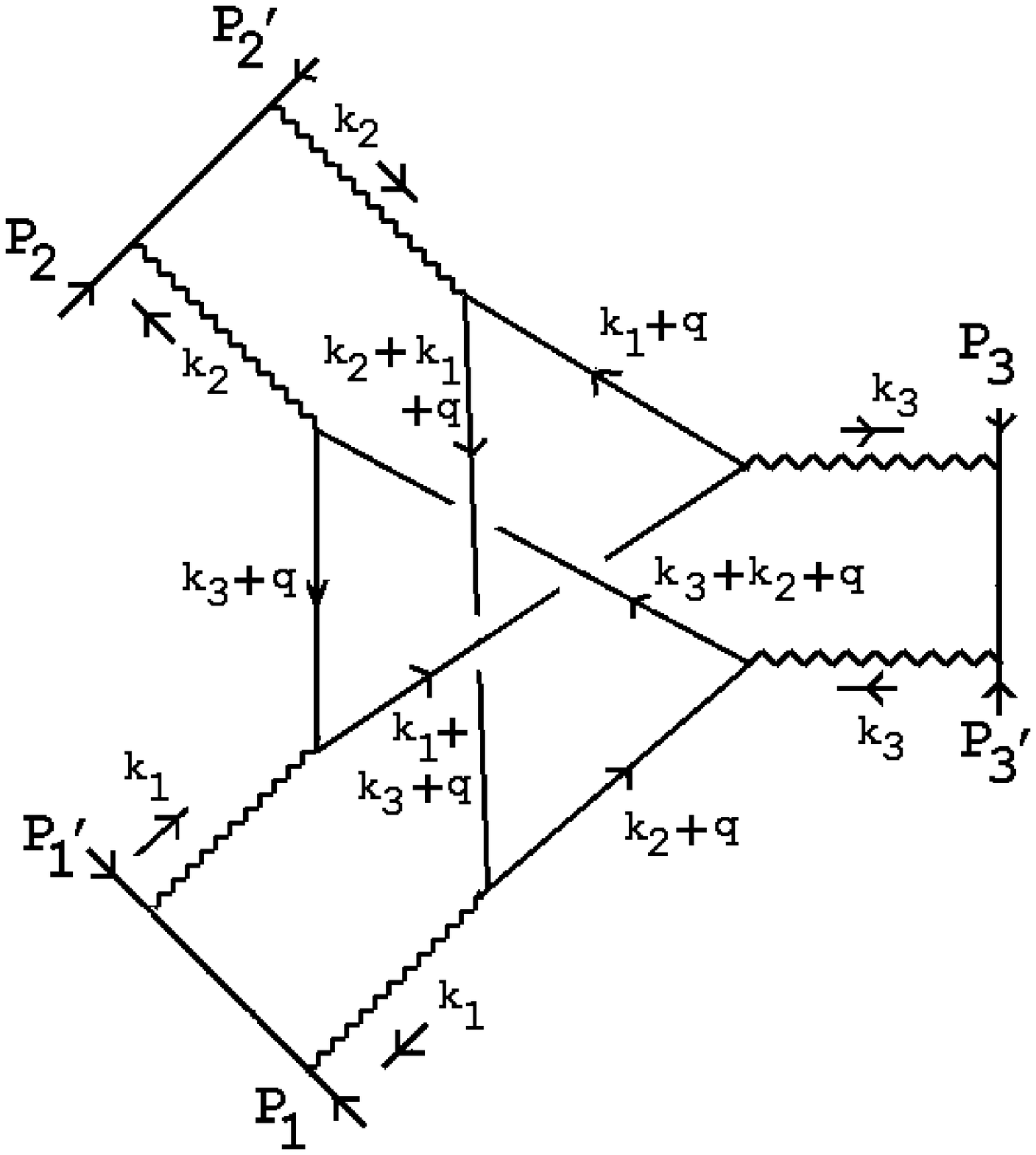}
\newline (b)
\end{center}}
\newline \centerline{Fig.~5.1 Routing Loop Momenta for Fig.~3.2.}

Using the momentum routing of Fig.~5.1(a)
and the analysis of the previous Section
we consider logarithms generated by the $k_1$ and $k_2$ 
integrations.
The $k_1$ and $k_2$ loops are shown in Fig.~5.2.
\begin{center}
\epsfxsize=4in
\epsffile{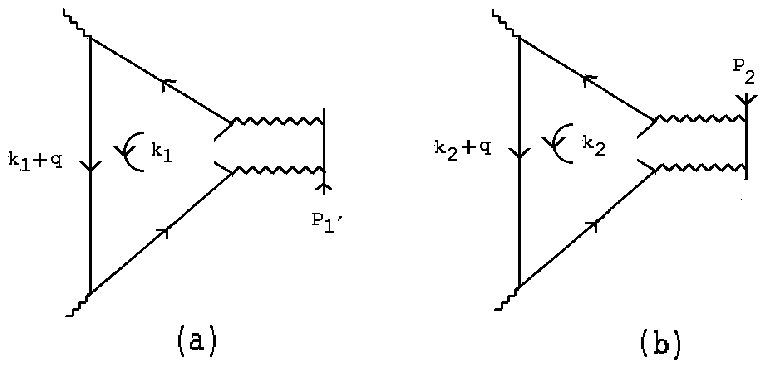}

Fig.~5.2 (a) The $k_1$ Loop (b) The $k_2$ Loop.
\end{center}
For the moment, we omit the propagators in the sloping lines
and all propagator numerators. (The omitted propagators will, nevertheless,
play an important role below. They are also
relevant if we wish to consider the other kinds of discontinuities
that appear in Fig.~3.3.)
In this case, the two loops differ only in the light-cone direction of
$P_1'$ and $P_2$.

We consider Fig.~5.2(a) first. We 
can directly apply the discussion following (\ref{lcan4})
if we identify $P_{1'}$ with $p$, 
$q$ with $p'$, 
$k_1$ with $k$, and consider the propagator pole at 
$(k_1+q)^2 = 0$. We then obtain 
$$
\eqalign{ I(p_{1'}q_{1^-})~\sim~&i\int 
d^2\underline{k}_{1\perp} \left[ -k^2_{1\perp} + 
i\epsilon \right]^{-2} ~\int_0^{\lambda q_{1^-}} dk_{1^-}  
\left[k_{1^-} -q_{1^-} \right]^{-1}\left[ p_{1'} k_{1^-}
- \underline{k}^2_{1\perp}
+i\epsilon \right]^{-1} \cr 
 \sim~&
{1 \over p_{1'}q_{1^-}}~\log{[p_{1'}\lambda q_{1^-} + i\epsilon]} } 
\auto\label{l50}
$$
We have used the notation (used extensively in the following) that for any
four-momentum $k$
$$
k_{i^-}~=~k_0 - k_i ~~~~~\underline{k}_{i\perp}~=~(k_j,k_k)~~j\neq k \neq i
~~~~~ ~~i,j,k~ = 1,2,3 
\auto\label{not-}
$$
The $q_{1^-}$ 
dependence indicates that the logarithm is a reflection of a threshold
in the invariant $P_{1'}.q$ . This dependence plays an important 
role in the following discussion. We also retain the $\lambda$-dependence, 
for technical reasons that will become apparent later. The final
result will be independent of $\lambda$, as it must be. 
From Fig.~5.2(b) we analagously obtain
$$
I(p_2q_{2^-})~\sim~~
{1 \over p_{2}q_{2^-}}~\log{[- p_{2}\lambda q_{2^-} + i\epsilon]}
\auto\label{l51}
$$
The minus sign appears relative to (\ref{l50}) because of the 
opposite direction of $P_2$.

Next we consider how the logarithmic branch cuts
generated by the  $k_1$ and $k_2$ integrations can trap the internal loop
integration over $q$ to produce an overall
discontinuity in $s_{1'2} \sim p_{1'}p_{2}$. 
For simplicity, we consider the region where
$$
\underline{k}_{i \perp}^2 ~\sim~ q^2~~\sim ~ 0 ~~~~~i~=~1,2,3
\auto\label{5an0}
$$
Appealing to (\ref{5an}) we can then, for our present purposes, effectively 
ignore the remaining 
$k_i$ dependence of the quark loop (including the propagators that
we ignored in the above discussion). If we parameterize $q$ as 
$$
q~=~\biggl(q_0,~q_{1^-},~q_{2^-},~q_{3^-} \biggr)
\auto\label{l52}
$$
we can treat the $q_{i^-}$ as independent variables, 
with $q_0$ essentially determined by the constraint $q^2 \sim 0$.
The logarithmic cuts of (\ref{l50}) and (\ref{l51}) appear, respectively,
in the $q_{1^-}$ and $q_{2^-}$ planes and if we make a further change of 
variables to
$$
q_{1^-}~=~x_2x_3~~, ~~~~q_{2^-}~=~x_3x_1 ~ ~, ~~~~~q_{3^-}~=~x_1x_2
\auto\label{l53}
$$
the two branch points appear in the $x_3$-plane, for fixed, positive, 
$x_1,x_2$, 
as illustrated in Fig.~5.3(a).
\begin{center}
\epsfxsize=5.8in
\epsffile{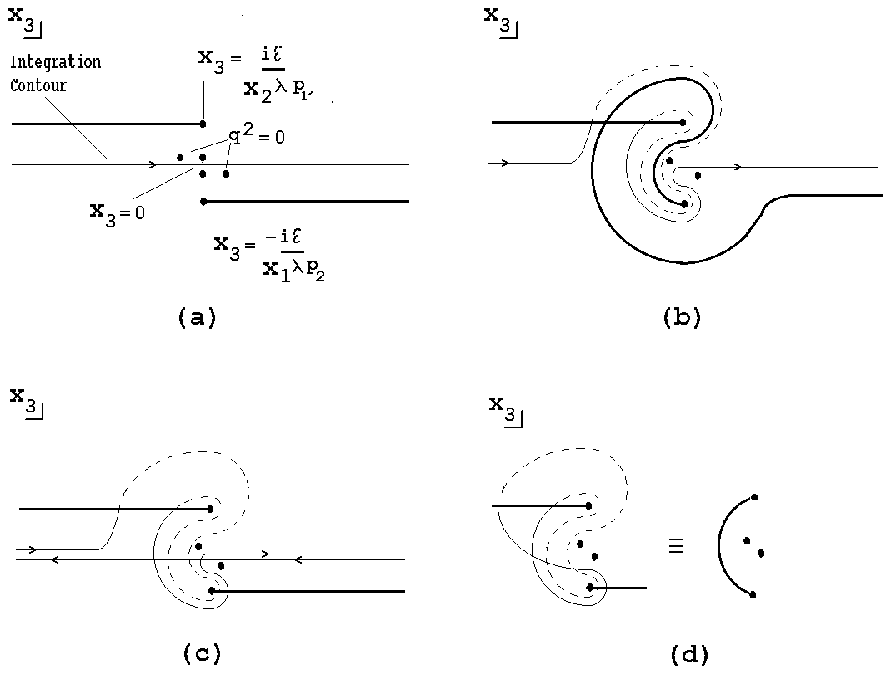}
\newline Fig.~5.3 Contours in the $x_3$-plane (a) the initial contour 
(b) $p_{2} \to e^{2\pi i}p_{2}$ 
\newline (c)  the discontinuity   
(d) the discontinuity as a line integral. 
\end{center} 
(The branch points also appear, separately, 
in the $x_2$ and $x_1$ planes. To focus on the
$s_{1'2}$ discontinuity and avoid any complication
from discontinuities involving a
logarithm of $p_3$ in these planes 
we can take the $\lambda$ for this logarithm to be much smaller.) 
The propagator poles that are not on-shell, that we ignored in the 
above discussion, combine to give a multiple pole at
$q^2=0$ (on both sides of the contour, as determined by
the presence of $i\epsilon$ in all propagators). 
If we continue to ignore propagator
numerators then the factors of $1/q_{1^-}$ and  $1/q_{2^-}$, 
in (\ref{l50}) and (\ref{l51}) respectively, will also contribute poles at 
$x_3=0$ (that will partly be compensated by the jacobian due to 
the change of variables).
However, in the anomaly contribution we will ultimately consider, these poles 
will be directly canceled by numerator factors.

The threshold we are interested in occurs when the two branch points collide
(at $x_3 = 0$ for $\epsilon = 0$). To extract the discontinuity we
consider a full-plane rotation of $p_{2}$, with $p_{1'}$ fixed, so that  
the logarithmic branch-cut
(\ref{l51}) deforms the contour as shown in Fig.~5.3(b) 
- the dashed line indicates that the contour is on the 
second sheet of the branch-point (\ref{l50}). (We have omitted the poles at
$x_3=0$.) Note that 
the continuation path we have chosen isolates the discontinuity
around the $s_{1'2}$ branch cut, since it avoids the pinching of the 
integration contour with the singularity at 
$q^2=0$ that would give other discontinuities. 
The desired discontinuity is obtained by adding 
the original contour in the opposite direction, as shown in Fig.~5.3(c).
Combining both contours 
we obtain Fig.~5.3(d) which, as illustrated can be written as a line integral 
between the two branch 
points of the double-discontinuity due to both cuts. As $\epsilon \to 0$, 
or in the asymptotic
limit $p_{1'}, p_2 \to \infty$, the branch points approach each other
and the result is a closed contour integral around the 
singularity at $q^2=0$ which is independent of the position of the end points 
and remains finite in the asymptotic limit. This is the asymptotic 
discontinuity and the singularity at $q^2 = 0$ 
is clearly crucial in producing a non-zero result. 

In Fig.~5.4(a) we have illustrated
the effect of adding 
(external and internal) transverse momenta in the 
the foregoing analysis. 
The integral between the branch points, of the double discontinuity,
is still obtained, while the 
singularity at $q^2 =0$ separates into a set of poles at both 
positive and negative $x_3$.
\begin{center}
\epsfxsize=5in
\epsffile{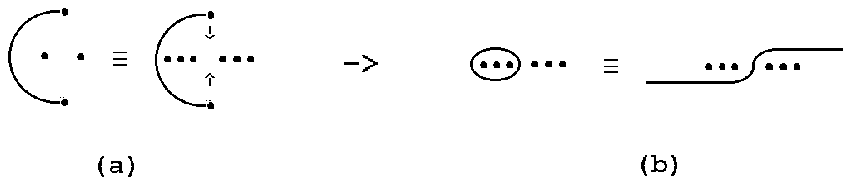}

Fig.~5.4 (a) The $x_3$
contour with finite transverse momenta (b) equivalence of the
asymptotic contour to the original contour.  
\end{center} 
In Fig.~5.4(b) we have  shown the asymptotic discontinuity.
Since the branch points are
logarithmic, the double discontinuity involved is simply $4 \pi^2$ and so 
no longer contains either branch cut.
Consequently, the asymptotically finite integral around 
the poles to the left can be opened up to give the original contour, 
as illustrated. (If there is a singularity at $x_3=0$, the contour is 
constrained to pass through this point although, as we noted above, for the
anomaly contribution to graphs, this will not be the case). 
The final result shown in Fig.~5.4(b) is just what
would be given by the normal cutting rules for a discontinuity in $s_{1'2}$ ,
i.e. the original integral with the four propagators involved in generating 
the discontinuity placed on-shell. Note that the same result 
is obtained if the discontinuity is evaluated by 
varying $p_{1'}$. An integral around the positive $x_3$ poles appears at the
intermediate stage, which can then 
be opened up to give the same final contour as in Fig.~5.4(b).  

An obvious, but essential, requirement in the
origin of the asymptotic discontinuity, which we want to emphasize, 
is that the branch-cuts due to the logarithms
in $p_{1'}$ and $p_2$ must lie on opposite sides of the $x_3$ contour.
In a physical region this requirement 
is normally straightforward for a loop integration producing a threshold
due to two massive states since the loop momentum will flow oppositely
through the two states and the $i\epsilon$ prescription will place
the states on opposite sides of the energy integration contour. In the 
variables we are using the generation of the threshold is a little more 
subtle.
Note, for example, that when $x_1 < 0$ the branch-point (\ref{l50}) appears 
in the upper half-plane (moving through infinity as $x_1$ moves through zero)
and there is no discontinuity. Therefore, the signs of the $x_{i}$ play an 
essential role in the occurrence of the discontinuity. 
A further requirement, which clearly holds in the case just discussed, 
is that the trapping (pinching)
of the contour that we have discussed 
must combine with the pinching associated with the logarithms
to give a complete cut through the diagram. That is to say, the complete set of
pinchings must correspond to an overall invariant cut.

We consider next the unphysical discontinuities that are our 
principal interest.
According to the discussion in Section 3, we are looking for a triple 
discontinuity of the form of Fig.~3.4 that treats the three cut lines of the
quark loop symmetrically so that, in a  
physical region, the sign of the energy component can be 
the same for all three on-shell states. We will, therefore, confine 
our discussion to a search for a symmetric triple discontinuity. 
As we noted, if the normal cutting rules apply 
there is no triple discontinuity (symmetric or not) of the Fig.~3.4  kind.
We consider whether the direct evaluation of discontinuities 
gives the same result. 

The discontinuity we 
discussed above occurred in a physical region that is unsymmetric 
in that $P_2$ is the momentum of an incoming particle while 
$P_1$ is the momentum of an outgoing particle.
To look for a symmetric discontinuity we will 
use an analysis that treats the complete graph symmetrically throughout.
To this end, we will start in the symmetric asymptotic region (\ref{np3})
where all momenta are real and 
$$
s_{i'j}~\sim ~-p_i p_j ~~<~0
\auto\label{ninv}
$$
In this region, the diagram is defined by the usual $i\epsilon$ prescription.
Since all three invariants must be positive, the triple discontinuity of
Fig.~3.4 can only be present in the triple-regge limit if we allow the large
momenta involved to be unphysical. A symmetric way to do this is to start 
from the real physical region and take
$$
p_i ~\to ~ e^{-i\pi /2} p_i~= i p_i~,~~i=1,2,3 ~~~~~
=> s_{i'j}~\sim ~ (- ip_i)(i p_j) ~~>~0  
\auto\label{pinv}
$$

Given the symmetry of the present discussion, 
it is immediately apparent that there will not be a (symmetric)
triple discontinuity, as we now show.
Using the above analysis, logarithms will be generated by each of the $k_i$ 
integrations. If we consider again the region where the 
transverse momenta are close to zero then, from (\ref{5an}), the requirement 
that the energy component of each on-shell line in the loop have the same sign
is equivalent to requiring that the $q_{i^-}$ all have the same sign. 
This, in turn, requires 
that the $x_i$ should all have the same sign.
However, in the symmetric real physical region, 
if $x_1$ and $x_2$ have the same sign, 
the logarithmic branch cuts in $P_1$ and $P_2$ lie on the same side 
of the $x_3$ contour as illustrated in Fig.~5.5.
\begin{center}
\epsfxsize=2in
\epsffile{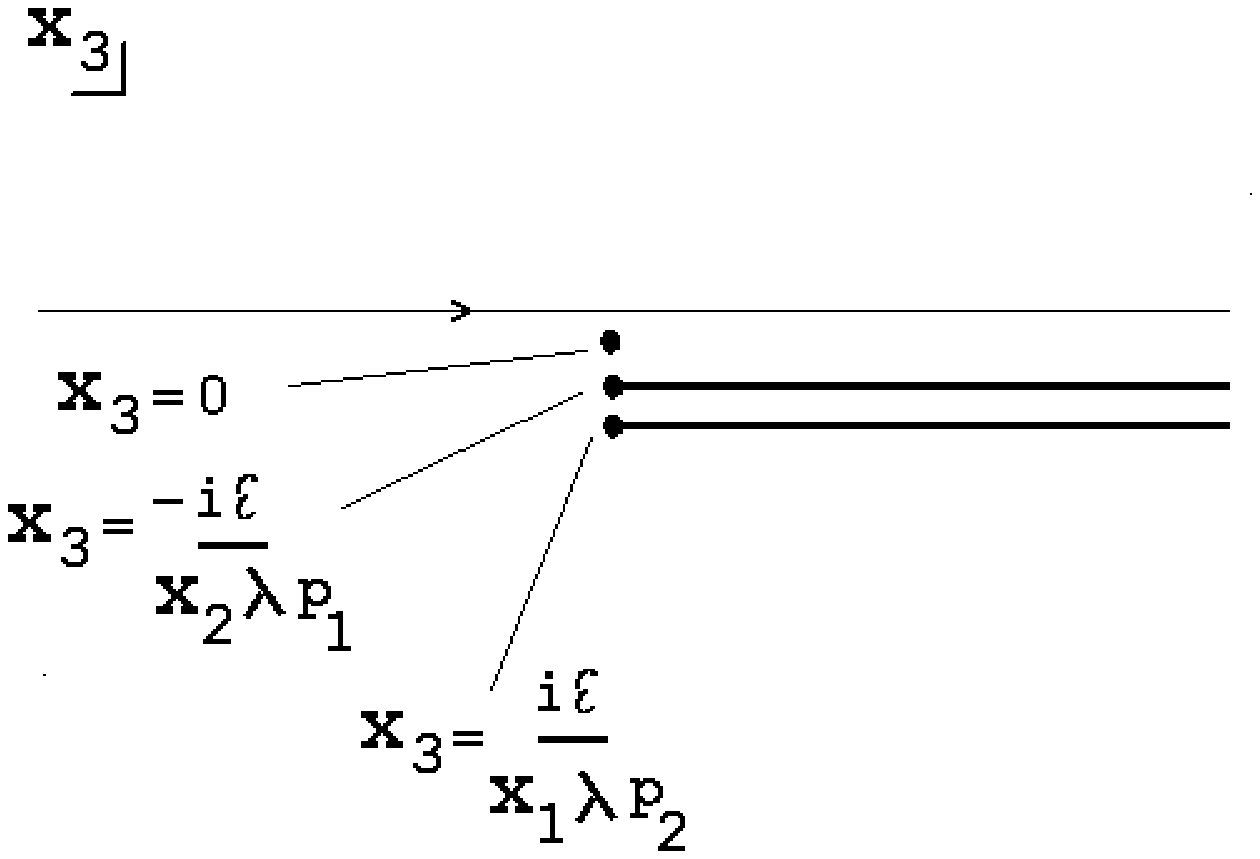}

Fig.~5.5 The Symmetric Location of Branch-Cuts in the $x_3$-plane. 
\end{center}
Since the continuation
(\ref{pinv}) is symmetric they will remain on the same side after the
continuation. 
As a consequence, in the symmetric $x_i$ region, 
the contour will not be trapped and distorted as 
one branch point moves aound the other, as it was in Fig.~5.3, and no
discontinuity will result. We conclude therefore that, 
for the graph we are discussing, 
discontinuities can only be generated in asymmetric 
regions of the $x_i$ that can 
not provide the symmetric triple discontinuity that we are looking for.

The foregoing analysis also precludes the occurence of a triple discontinuity,
that is appropriately symmetric, in the diagram of Fig.~3.5. 
To obtain a symmetric triple discontinuity we look for a graph that has the
appropriate overall symmetry and also, for each $i \neq j \neq k$,
has logarithmic branch
cuts on both sides of the $x_i$ contour
in a symmetric region of $x_j$ and $x_k$. With these requirements in mind, 
an obvious graph to consider is that of Fig.~3.7. To discuss this graph we 
continue, for simplicity, to take $Q_1=Q_2=Q_3=0$. Two symmetric (distinct)
routes for the internal momenta are shown in Fig.~5.6. 
For a threshold corresponding to the cutting of particular
lines of the internal quark loop to be generated the  
external loop momentum generating the relevant logarithms
must pass through at least one of the lines. With this constraint,
only the routing shown in 
Fig.~5.6(a) will give both discontinuities of the kind we are looking for
and the $\gamma$-matrix structure for on-shell contributions
that we show, in the next Section, gives the anomaly.
The routing of Fig.~5.6(b) 
would be appropriate for discussing 
the triple discontinuity of Fig.~3.6. However, in this case the 
$\gamma$-matrix structure needed to generate the anomaly does not 
appear in the on-shell contributions. Therefore, the triple discontinuity 
of Fig.~3.6 does not contain the anomaly.
\newline \parbox{3in}{ 
\begin{center}
\epsfxsize=2.4in
\epsffile{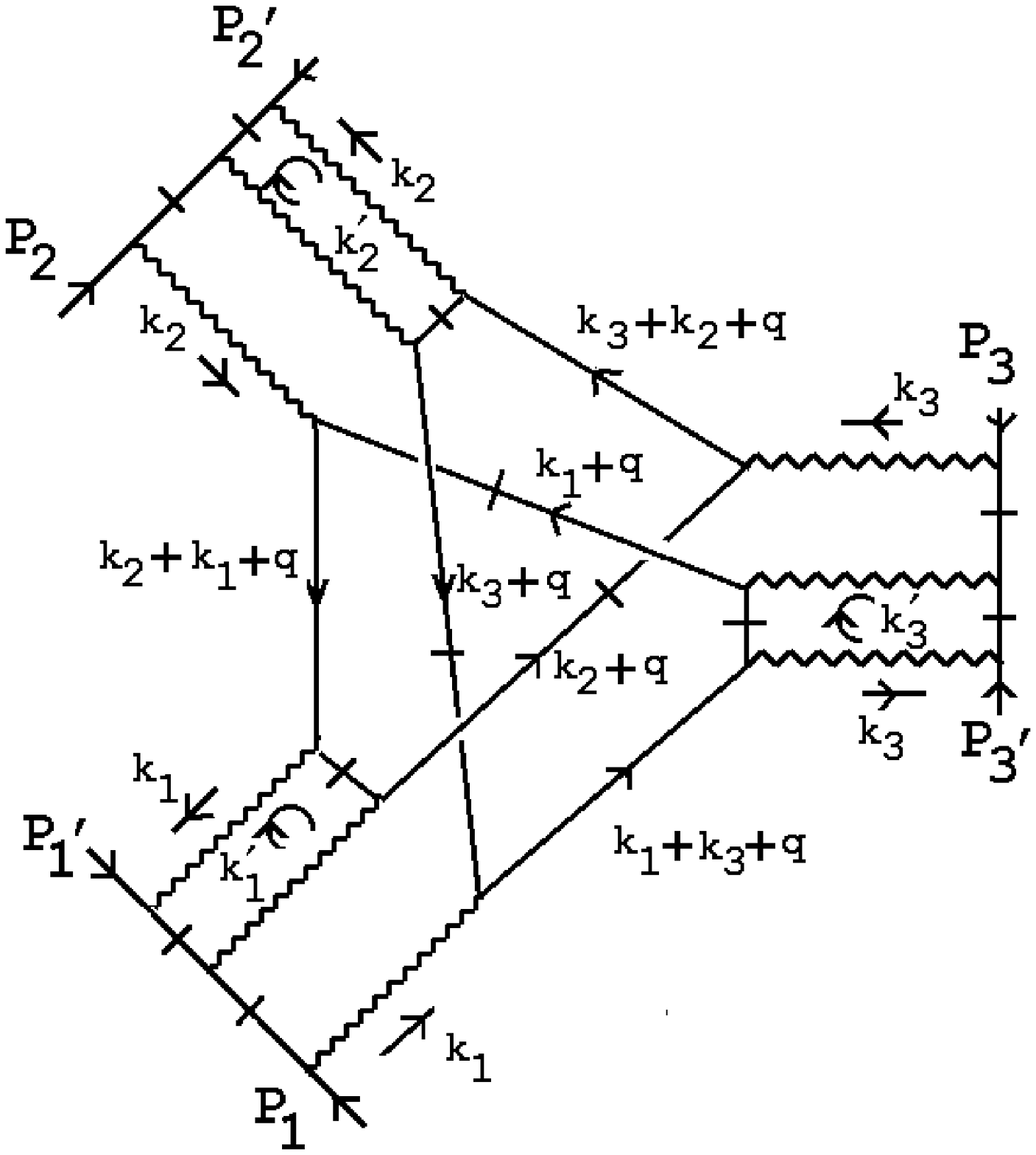}
\newline (a)
\end{center}}
\parbox{3in}{ 
\begin{center}
\epsfxsize=2.4in
\epsffile{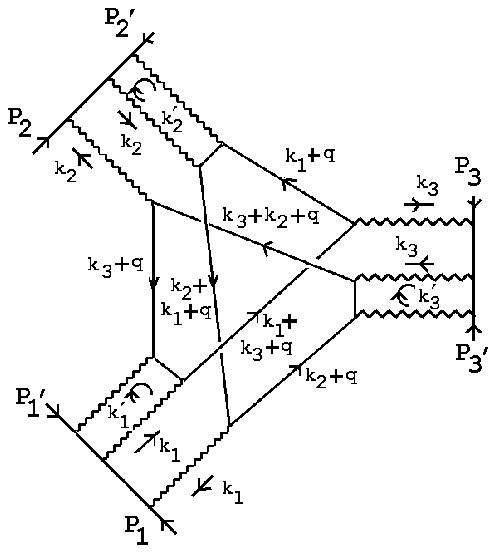}
\newline (b)
\end{center}}
\begin{center}
Fig.~5.6 Labeling Momenta for Fig.~3.7.
\end{center}

Using the momentum routing of Fig.~5.6(a) 
we consider the logarithms generated by both the $k_i$ and $k_i'$ loop 
integrations. Extracting all logarithms places on-shell all the hatched lines
of Fig.~5.6(a), and 
gives leading behavior of the form of (\ref{211}) multiplied by double 
logarithms of each of the $P_{i^+}$.
At the diagrammatic level (i.e. temporarily 
discussing diagrams rather than discontinuities), we anticipate that
existing calculations can be adapted to show that 
the double logs are canceled by adding diagrams of the kind illustrated
in Fig.~5.7. That is, we add diagrams containing twists relative to Fig.~3.7,
as in Figs.~5.7(a) and (b) together with diagrams, 
such as that in Fig.~5.7(c), that 
produce the well-known cancelations necessary for reggeization.
\begin{center}
\epsfxsize=5in
\epsffile{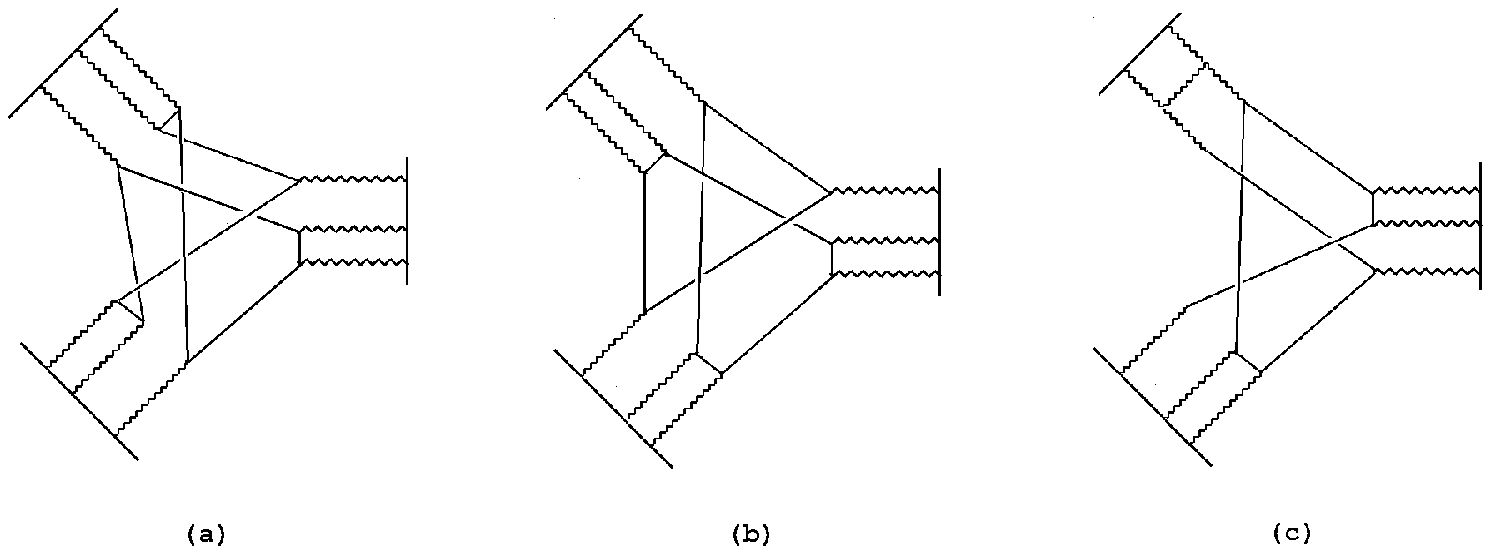}
\newline Fig.~5.7 Diagrams with (a) Twisted $k_i'$ Loops (b) Twisted $k_i$ 
Loops 
(c) Reggeization Cancelations
\end{center}

After the double logs 
are canceled, the remaining single logs should go into 
reggeization contributions, in analogy with Fig.~3.6, with 
the remaining terms providing new lowest-order
reggeon interactions. As we have emphasized repeatedly, to discuss this 
systematically we consider multiple asymptotic discontinuities 
rather than the behavior of full diagrams. We do this, as above, 
by keeping the $q$-dependence of all logarithms together with all
$i\epsilon$ dependence.
We consider specifically the logarithms generated by
the $k_1$ and $k_1'$ loops, but the symmetry of the diagram 
obviously determines that the others can be treated identically. The loops,
extracted from Fig.~5.6, are shown in Fig.~5.8.
The $k_1$ loop is identical to those of Fig.5.2 and 
can be evaluated analagously.
\begin{center}
\epsfxsize=3in
\epsffile{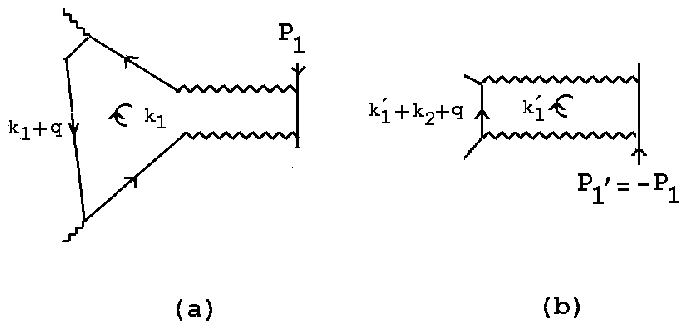}
\newline Fig.~5.8 (a) The $k_1$ Loop (b) The $k_1'$ Loop.
\end{center}
Using a similar analysis, the $k_1'$ loop gives an integral of the form
$$
\int_0^{(k_2+q)_{1^-}}~dk_{11^-}' ~~\cdots
\auto\label{l54}
$$

If we again go to the region where all transverse momenta are close 
to zero then,
using (\ref{5an}), it follows that after the $k_2$ integration
$$
k_{21^-} ~ \sim ~k_{20} ~\sim ~q^2/q_{2^-}~ << q_{1^-}
\auto\label{l55}
$$
Therefore, we can take the upper end-point in (\ref{l54}) to be $q_{1^-}$. In 
this case both the $k_1$ and $k_1'$ integrations give logarithms with $q_{1^-}$
in the argument - but with opposite signs.
We then have branch-cuts located as in Fig.~5.9(a) in each of the 
$x_1,x_2$ and $x_3$ planes.
We have included poles at $q^2=0$ and $x_i=0$
and have used different $\lambda_i$ and $\lambda_i'$
for each branch-cut to allow us to separate the branch points in our 
discussion. 

With values of the $\lambda_i$ and $\lambda_i'$ implied 
by Fig.~5.9(a),
we could clearly obtain a discontinuity in $s_{jk'}$ (due to the two
closest branch points) by repeating the 
discussion illustrated by Fig.~5.3. The discontinuity would similarly be an 
integral between the two branch points involved, as in Fig.~5.3(d), 
but because of the additional
branch points that are present, the contour could not be opened up as in 
Fig.~5.4. Therefore, having taken $x_j, x_k > 0$ so that the branch 
cuts lie  as in Fig.~5.9(a),
the discontinuity would involve only pure imaginary or negative 
real part values of $x_i$.
\begin{center}
\epsfxsize=5.5in
\epsffile{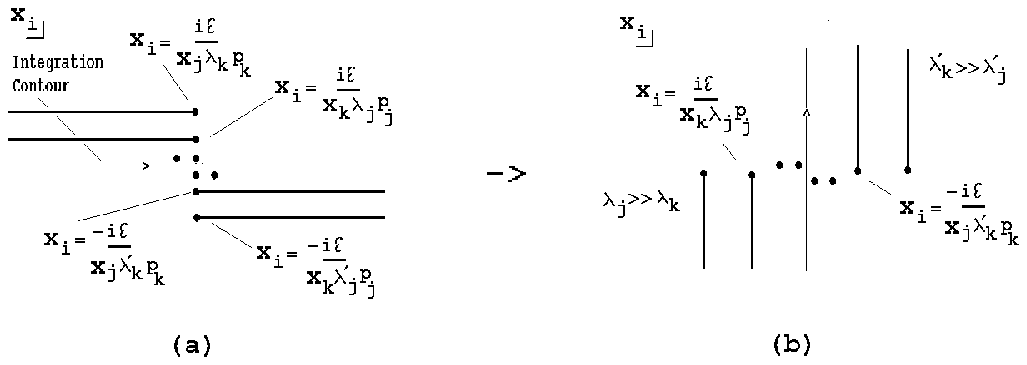}
\newline Fig.~5.9 (a) Branch Points in the $x_i$-plane 
(b) $p_i ~\to ~ e^{-i\pi /2} p_i~= i p_i~,~~i=1,2,3$ 
\end{center}
 Consequently,
any further discontinuity obtained by the collision of 
branch points in the $x_j$ or $x_k$ planes would have to involve mixed 
real part signs for
the $x_i$. We conclude (not surprisingly) that 
in the physical region a triple discontinuity can not be obtained that 
involves only positive values of all three $x_i$.

This brings us to the central point of the paper. If
we go to the unphysical
region (\ref{pinv}), where we expect to encounter
an unphysical triple discontinuity, the last analysis changes in a crucial
manner. The 
resulting location of branch cuts is now as shown in Fig.~5.9(b), allowing
the integration contour to be rotated as illustrated. 
In Fig.~5.9(b) we have also, for emphasis, 
chosen significantly different values of the $\lambda_i$ and $\lambda_i'$.
If we again determine the discontinuity associated with the collision of the 
two nearest branch points, as above, the result will be the contour 
integral of the double discontinuity shown in Fig.~5.10.
\begin{center}
\epsfxsize=4in
\epsffile{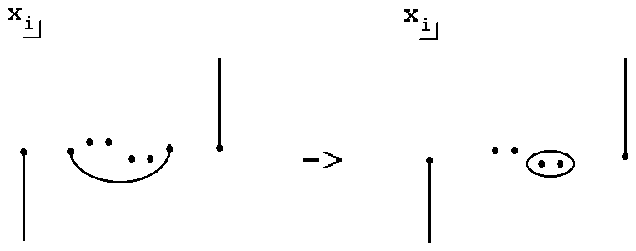}

Fig.~5.10 The unphysical region discontinuity.
\end{center}
Now the integral involves positive real values of $x_i$ and, as illustrated,
the asymptotic limit gives a loop integral over just positive values.
The contour integral can not be opened up, however, since the other branch cuts
remain.

Having derived a first discontinuity from two branch points in the 
$x_i$ plane, as in Fig.~5.10, it is straightforward
to keep the remaining branch points and move on to the $x_j$ and $x_k$ planes
where, in each case, only two branch cuts now appear. In both planes,
discontinuities of the form of Fig.~5.10 occur, provided the 
$x_i$ integration is restricted to positive real values. Therefore, we obtain 
a triple discontinuity in which each of the $x_i$, $x_j$ and $x_k$
integrations is consistently over positive values and the asymptotic 
contour is obtained as illustrated by the first two contours in Fig.~5.11.
\begin{center}
\epsfxsize=4in
\epsffile{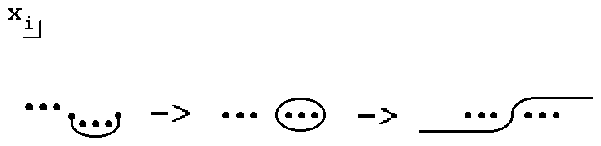}

Fig.~5.11 Contours for the $x_i$, $x_j$ and $x_k$ integrations.
\end{center}
Since all logarithmic branch cuts are now removed, all three contours can 
be opened up to obtain the last contour of Fig.~5.11 which is, once again the
original contour of integration for each of $x_i$, $x_j$ and $x_k$.
We thus obtain a triple discontinuity which, at firat sight, corresponds to
the usual cutting rules since all cut lines are on-shell. However, there 
is a subtlety.

If we consider the discontinuity arising from the pinching of logarithms 
of $p_1\lambda_1$ and $p_2\lambda_2'$, for example, then the lines  
put on-shell in the discontinuity are those that have thick hatches in
Fig.~5.12(a).
\begin{center}
\epsfxsize=4.5in
\epsffile{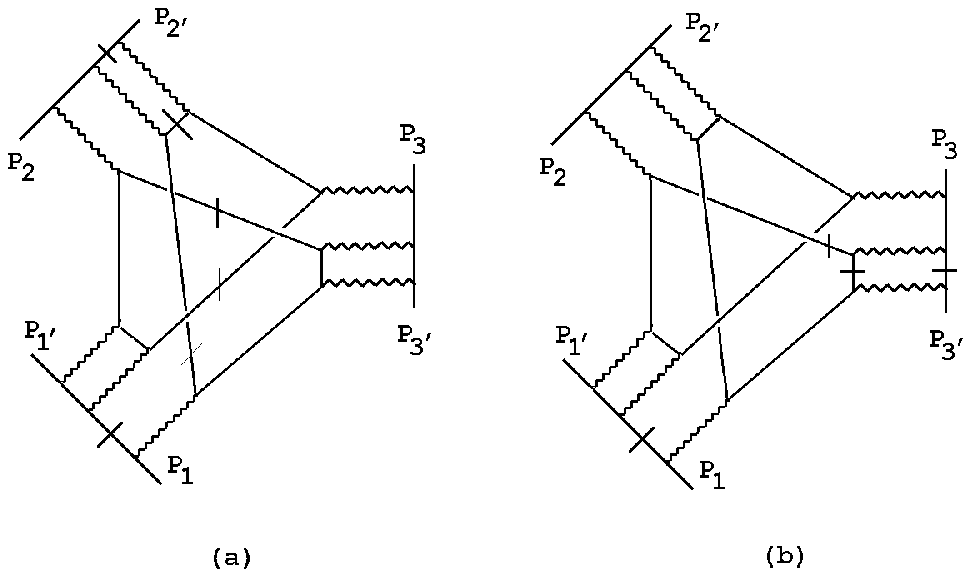}

Fig.~5.12 On-shell lines for (a) an $s_{12'}$ discontinuity (b) a potential
$s_{13'}$ discontinuity.
\end{center}
These lines are only a subset of those required to obtain a 
complete cut of the diagram. This implies that the corresponding
pinching does not, by itself, give a singularity of the complete integral and
a-priori the integration contour could be deformed away from the pinched
region.
To obtain a complete cut we must add the lines that have thin hatches in
Fig.~5.12(a). When 
these lines are on shell the pinching does give an overall singularity.
But, if we require a common sign for the $x_i$
the two thin-hatched lines actually have the wrong $i\epsilon$ 
prescription to straightforwardly combine with the asymptotic
pinching to give what would be a physical sheet 
``asymptotic normal threshold''. However,
each of the two thin hatched lines is separately placed on shell 
by one of the additional discontinuities. Therefore, a triple discontinuity
of the kind we have found does correspond to the triplet 
$\{s_{12'},s_{23'},s_{32'}\}$ of invariant cuts. 

Note that if we consider instead 
the discontinuity arising from the pinching of logarithms 
of $p_1\lambda_1$ and $p_3\lambda_3'$ then the lines put on shell are 
those hatched in Fig.~5.12(b). In this case there is no simple way to 
include additional lines and obtain an invariant cut. Therefore, this pinching
can not be extended to a complete cut of the diagram.
We conclude that the triple discontinuity in $\{s_{12'},s_{23'},s_{32'}\}$
that is illustrated in Fig.~3.7 is the only combination that exists, 
as an extension of the above analysis. It
is symmetric, with each of the internal quark lines 
that are put on shell by $k_i$ integrations treated symmetrically. 
All three of these lines contribute to each invariant cut but, as we 
have just discussed, two of them always have the 
wrong $i\epsilon$ prescription, relative to the third,
to give a physical normal threshold.
Singularities associated with combinations of forward and backward 
going particles
(as the mixture of $i\epsilon$ prescriptions implies is the case) are
``mixed-$\alpha$'' solutions of the Landau equations\cite{arw00} and
are referred to as pseudothresholds. In general, pseudothresholds are not 
singular on the physical sheet, just because of the conflicting 
$i\epsilon$ prescriptions. However, they are generally singular on 
unphysical sheets and can appear in multiple 
discontinuities. For the unphysical multiple 
discontinuity we are discussing, a combination of 
``asymptotic pseudothresholds'' can contribute when the same combination of 
normal thresholds can not.

\newpage

\mainhead{6. THE TRIANGLE ANOMALY }

In this Section we give a brief discussion of how the anomaly occurs in the
triple discontinuity of Fig.~3.7. A complete discussion would be obtained 
by a straightforward generalization, to include the minor
additional complexities, of
the lengthy analysis of Fig.~3.1 in \cite{arw99}.    

All the cut lines of Fig.~3.7 are on-shell, as described in the last Section.
We begin by 
adding in the numerator dependence that we essentially ignored in the 
previous Section. For the external lines, additional powers of the external
momenta are generated as in (\ref{lcan61}) and (\ref{coup}). As a result,
inverse external momentum factors, such as ${p_{1'}}^{-1}$ in (\ref{l50}) and 
 ${p_2}^{-1}$ in (\ref{l51}) are eliminated and the 
factor of $P_{1^+}P_{2^+}P_{3^+}$ that appears in (\ref{211}) is produced.
Also, if we use the natural transverse momenta
given by (\ref{not-}), the light-like 
$\gamma$-matrix couplings that appear at each of the vertices of the internal loop
(after the triple-regge limit is taken) are as illustrated in Fig.~6.1(a). 
For the hatched lines that appear in both
Fig.~6.1(a) and (b), we keep the $\gamma$ matrices shown. 
These are the ``local couplings'' (see \cite{arw99}) that appear when that 
part of the associated numerator is kept that cancels the 
internal momentum factors, such as ${q_{1^-}}^{-1}$ in (\ref{l50}) and 
 ${q_{2^-}}^{-1}$ in (\ref{l51}), that arise from the longitudinal
loop momentum integrations. 
\begin{center}
\epsfxsize=5.8in
\epsffile{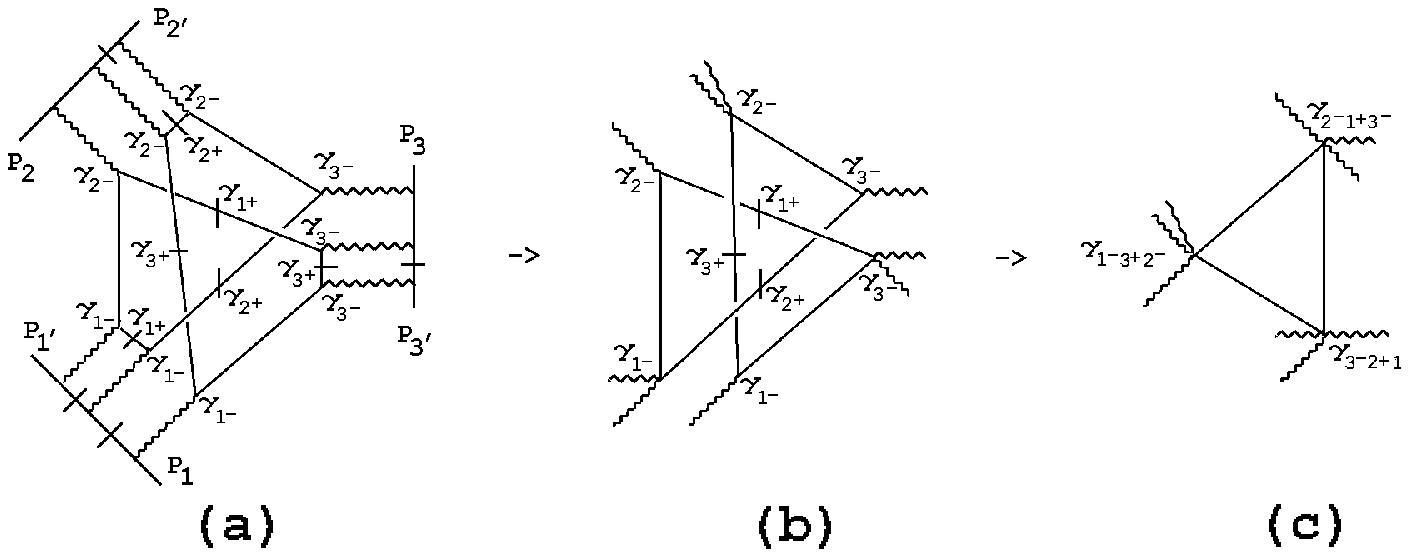}

Fig.~6.1 $\gamma$-matrix structure for the reggeon interaction extracted from
Fig.~3.7.
\end{center}
The resulting asymptotic behavior then has the form
$$
\eqalign{ ~~~~~P_{1^+}~ P_{2^+}~ P_{3^+}~
\prod_{i=1}^3 \int & { d^2 k_{i1}d^2 k_{i2} d^2 k_{i3}\over  
k_{ i1}^2  k_{i2}^2 k_{i3}^2 }  
~~ \delta^2 (Q_{i\perp} -  k_{i1} -  k_{i2}  - k_{i3})~G^3_i(k_{i1},k_{i2},k_{i3} 
\cdots) \cr 
&~~~~~~~~\times ~ R^9(Q_1,Q_2,Q_3,
k_{11}, k_{12},k_{13} \cdots )} \auto \label{611}
$$
where $R^9$ is the triangle diagram illustrated in Fig.~6.1(c).

By comparing with the three-reggeon version of (\ref{2ra1}) and (\ref{2ra})
we can extract $R^9$ as a nine-reggeon interaction which, if we now write 
$$
k_{i1} ~= ~q_i + k_i~, ~~~~ k_{i2} ~= ~q_i - k_i -k_i'~, ~~~~
k_{i3} ~= k_i'~,
\auto\label{dki6}
$$
can be written (very similarly to (\ref{580})) as 
$$
\eqalign{ &R^9(q_1,q_2,q_3,k_1,k_2,k_3,k_1',k_2',k_3') ~=\cr
& \int d^4 k  {  Tr \{ 
\gamma_5 \gamma^{1^-3^+2^-}(\st{k}+ \st{k}_1 + \st{q}_2 +\st{k}_3) 
\gamma_5 \gamma^{2^-1^+3^-} \st{k} 
\gamma_5 \gamma^{3^-2^+1^-}(\st{k}- \st{k}_2 + \st{q}_1 + \st{k}_3 )\} 
\over  (k + k_1 + q_2 + k_3 )^2  
~k^2 ~
 (k - k_2 + q_1 + k_3)^2 } }
\auto\label{612}
$$
where 
$$ 
\eqalign{\gamma^{1^-3^+2^-}~&=~
\gamma_{1^-}\gamma_{3^+}\gamma_{2^-} ~=~\gamma^{-,-,-}~-~ i~
\gamma^{-,-,+}  ~\gamma_5 \cr
\gamma^{2^-1^+3^-}~&=~\gamma_{2^-}\gamma_{1^+}\gamma_{3^-}
 ~=~\gamma^{-,-,-}~-~ i~
\gamma^{+,-,-}  ~\gamma_5 \cr
\gamma^{3^-2^+1^-}~&=~\gamma_{3^-}\gamma_{2^+}\gamma_{1^-}
 ~=~\gamma^{-,-,-}~-~ i~
\gamma^{-,+,-}  ~\gamma_5 }
\auto\label{g63}
$$
and $\gamma^{\pm,\pm,\pm}$ is defined by (\ref{g64}).

Because of the symmetric choice of co-ordinates and the completely symmetric
manner in which we 
have evaluated the triple discontinuity, the anomaly appears in a 
slightly different way to that discussed for Fig.~3.1 in \cite{arw99}.
To obtain the anomaly divergence
we must have a component of the axial-vector triangle diagram
tensor  $\Gamma^{\mu\nu\lambda}$  with
$\mu= \nu $ having a lightlike projection and $\lambda $ 
having an orthogonal spacelike projection. There must also be 
a transverse momentum (scaled to zero) in the remaining orthogonal
spacelike direction. If we choose the $\gamma_5$
component from all three vertices, 
the first requirement is not met. However, if we choose the 
$\gamma_5$ component from one of the three 
vertices in Fig.~6.1(c), and choose the vector coupling from the other
two vertices, it is met.
The finite light-like momentum involved must then have a projection on 
$n^{-,-,- \mu}$ and the orthogonal spacelike momentum must be distinct
in each case.
The three distinct possibilities for the anomaly to occur are associated 
with the three distinct hexagraphs described in \cite{arw99}, 
and hence with three distinct 
helicity amplitudes. We will discuss this relationship further in our later 
papers.

As we discussed at length in \cite{arw99}, while the triple discontinuity
giving the interaction of Fig.~6.1 occurs in an unphysical region, 
the interaction will, nevertheless, provide a ``real'' reggeon interaction
in physical regions. Because the discontinuity has the symmetry property 
that we emphasized in previous Sections,
the anomaly infra-red divergence can occur in the physical-region
configuration shown in Fig.~6.2. (The large dots indicate that a local
interaction is involved.)
The $\gamma_5$ interaction is at the intermediate vertex and the 
light-like momenta are as in(\ref{chm1})-(\ref{chm30}).  
Fig.~6.2 can then be identified with
the basic anomaly process of Fig.~3.2 except
that, as anticipated in Section 3, there is an additional wee gluon involved.
There are also additional gluons with finite transverse momentum.
\begin{center}
\epsfxsize=2.5in
\epsffile{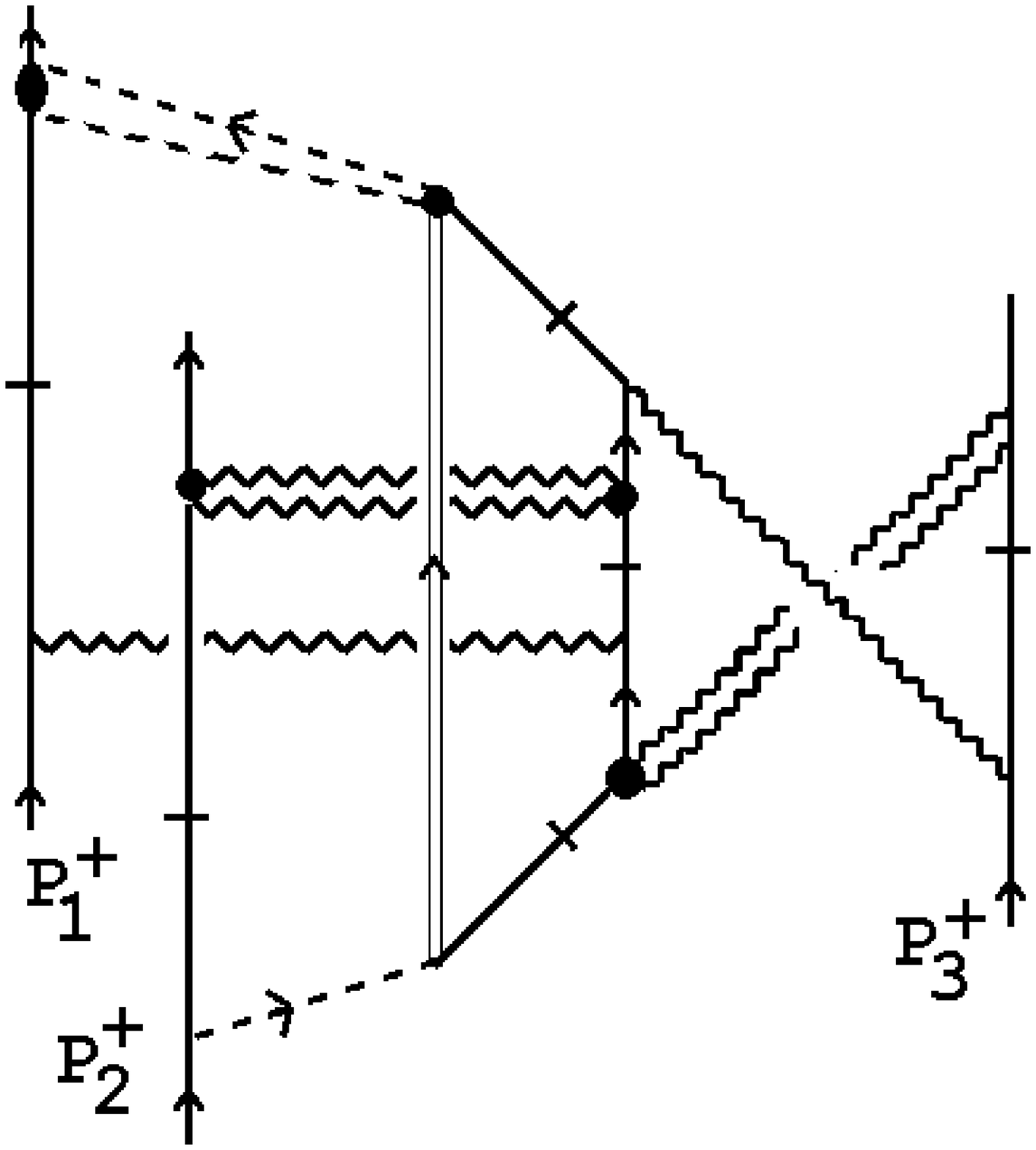}

Fig.~6.2 Physical region configuration for the anomaly divergence in Fig.~6.1.
\end{center}

If there are no
reggeization logarithms of the same order that appear accompanying the anomaly, 
as our discussion in Sections 3 and 5 implies, then that
part of the 
triple discontinuity interaction given by Fig.~6.1 that contains the anomaly
will appear as the leading triple-regge coupling of the three three-reggeon
states. All other diagrams that contribute will then have a similar triple 
discontinuity. The discussion in Section 5 shows that such diagrams must have 
``right-hand  and left-hand'' cuts in each $x_i$-plane, suggesting that only
diagrams having the same structure as that of Fig.~3.7, 
but with incoming and outgoing lines switched, can contribute. If this is the case,
signature conservation will occur, requiring an additional reggeon in at least one
channel. In this paper we will not introduce color factors, except to note that
we expect every reggeon state coupling to the anomaly
to carry anomalous color parity. This will ensure that the anomaly does not occur in
the scattering of elementary quarks and gluons - as we have anticipated.
Instead the scattering states must have an essential ``non-perturbative'' wee-parton
content that ensures they can scatter by 
exchanging reggeon states coupling to the anomaly.

We will postpone all further discussion of cancelations to 
our following papers. The purpose of this paper
has been to establish that a class of diagrams contain  
a multiple discontinuity that does generate the reggeon interaction anomaly. 
For the moment we note only that when the SU(3) gauge 
symmetry of QCD is broken to SU(2) the infra-red divergence that involves the
anomaly and that actually dominates bound-state interactions
occurs in diagrams that are very similar to the ones we have discussed.
An example, corresponding to a triple-regge multi-pomeron interaction,
is shown in Fig.~6.3.
\begin{center}
\epsfxsize=2.5in
\epsffile{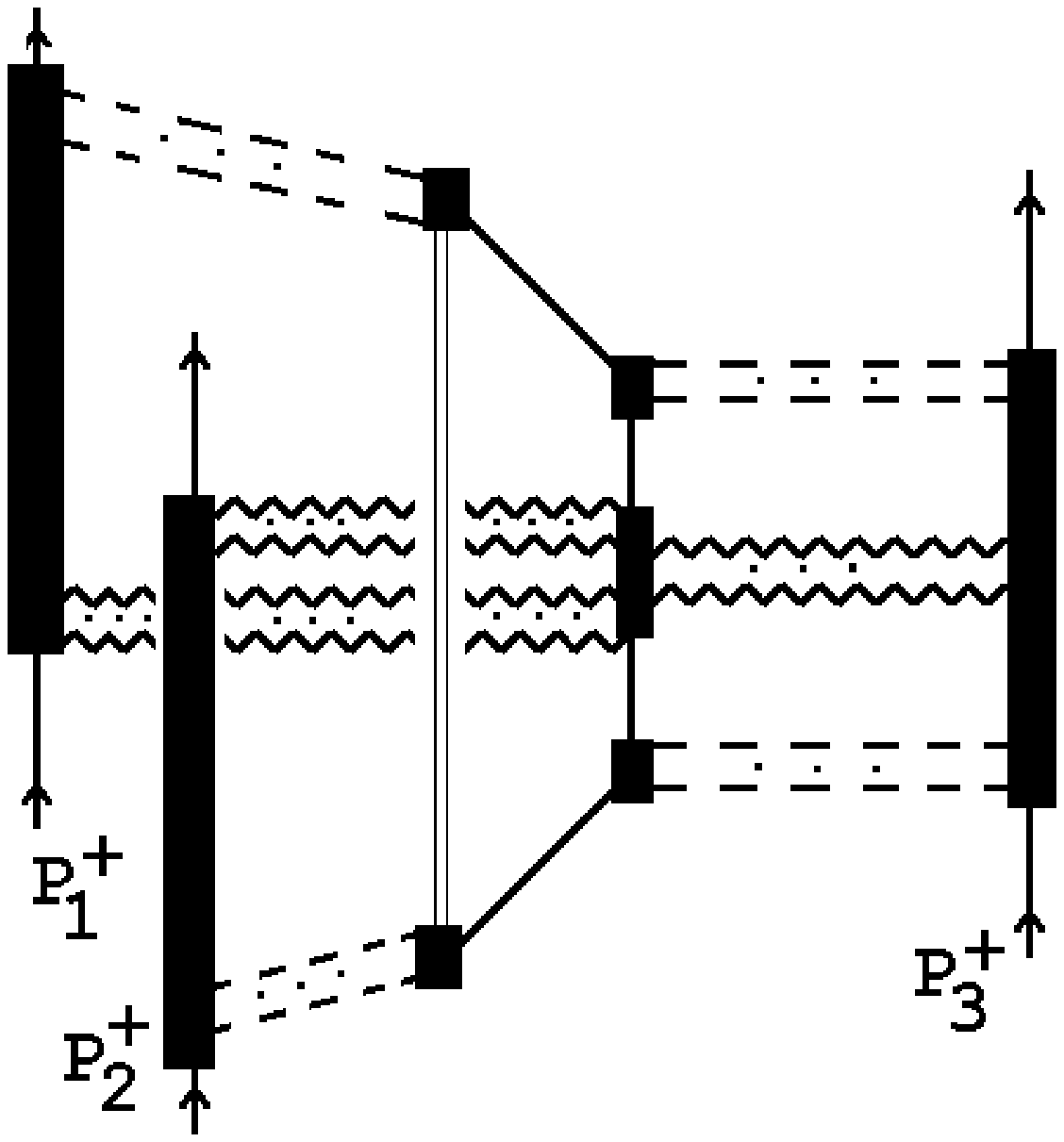}

Fig.~6.3 The anomaly configuration in bound state interactions. 
\end{center}
The scattering states are bound-states and the solid, wavy, lines
are, reggeized, massive gluon states that are SU(2) 
singlets. The dashed lines represent massless 
gluons carrying zero transverse momentum. In this situation the three
multi-reggeon (pomeron) states that are interacting through the anomaly 
all have a wee-parton component that participates in the divergence.

\end{document}